\documentclass[a4paper]{article}
\usepackage[a4paper, total={6in, 9in}]{geometry}
\usepackage{floatrow}
\usepackage{hyperref}
\usepackage{filecontents}
\usepackage{subfigure}
\usepackage{bm}
\usepackage[font={scriptsize,bf}]{caption}
\usepackage{amsmath}
\usepackage{amsfonts}
\usepackage{amssymb}
\usepackage{graphicx}
\usepackage{bm}
\usepackage{multicol}
\usepackage{multirow}
\usepackage{epstopdf}
\usepackage{soul}
\usepackage{pifont}
\usepackage{lineno}
\usepackage[table]{xcolor}
\usepackage{color}
\usepackage[autostyle]{csquotes}
\usepackage{tikz,pgf}
\usepackage{collcell}

\usepackage{pgfplots}
\usepackage{wasysym}
\usepackage{lipsum}
\usepackage{algorithm} 
\usepackage{algpseudocode} 
\newlength\fheight
\newlength\fwidth
\newcommand{\cmmnt}[1]{\ignorespaces}
\usepackage{etoolbox}
\usepackage{setspace}
\usepackage{csquotes}

\title{Efficient Transonic Aeroelastic Model Reduction Using Optimized Sparse Multi-Input Polynomial Functionals}
\author{ \large Michael Candon$^1$\thanks{corresponding author, candon.michael@rmit.edu.au}, Maciej Balajewicz$^2$, Arturo Delgado-Gutierrez$^1$, \\ \large Pier Marzocca$^1$, and Earl H. Dowell$^3$}

\date{
	\normalsize $^1$School of Engineering (Aerospace Engineering), RMIT University, Melbourne, AUS, 3000\\
	$^2$Independent Researcher, Boulder, CO, USA, 80304\\
 $^3$Duke University, Durham, NC, USA, 27708\\[2ex]%
}

\begin{document}
	
	\maketitle
	\begin{abstract}
        Nonlinear aeroelastic reduced-order models (ROMs) based on machine learning or artificial intelligence algorithms can be complex and computationally demanding to train, meaning that for practical aeroelastic applications, the conservative nature of linearization is often favored. Therefore, there is a requirement for novel nonlinear aeroelastic model reduction approaches that are accurate, simple and, most importantly, efficient to generate. This paper proposes a novel formulation for the identification of a compact multi-input Volterra series, where Orthogonal Matching Pursuit is used to obtain a set of optimally sparse nonlinear multi-input ROM coefficients from unsteady aerodynamic training data. The framework is exemplified using the Benchmark Supercritical Wing, considering; forced response, flutter and limit cycle oscillation. The simple and efficient Optimal Sparsity Multi-Input ROM (OSM-ROM) framework performs with high accuracy compared to the full-order aeroelastic model, requiring only a fraction of the tens-of-thousands of possible multi-input terms to be identified and allowing a 96$\%$ reduction in the number of training samples. 
        \end{abstract}

\section*{Nomenclature}

\begin{tabbing}
XXXXXXXXXX \= XX \= \kill
$b$ \>\> Semi-chord [m]\\
$C_L$ \>\> Lift coefficient\\
$C_M$ \>\> Pitching moment coefficient\\
$C_p$ \>\> Pressure coefficient\\
$\bm{D^p}$ \>\> Tensor of $p^{th}$-order Taylor partial derivatives\\
$\bm{d}$  \>\> Matrix of flattened tensors of linear, nonlinear and cross-terms \\
$\bm{d_s}$  \>\> Matrix of flattened sparse tensors of linear, nonlinear and cross-terms  \\
$\bm{F}$ \>\> Aerodynamic force vector (nodal coordinates)\\
$h$ \>\> Heave displacement [m]\\
$\mathcal{H}_0$ \>\> Steady-state generalized forces\\
$\bm{K}$ \>\> Stiffness matrix (nodal coordinates)\\
$\bm{GK}$ \>\> Generalized stiffness matrix\\
$\bm{L}$ \>\> Lower triangular circulant multi-input matrix\\
$m$ \>\> Number of structural modes\\
$\bm{\mathcal{M}}$ \>\> Nonlinear multi-input matrix\\
$\bm{M}$ \>\> Mass matrix (nodal coordinates)\\
$\bm{GM}$ \>\> Generalized mass matrix\\
$M_\infty$ \>\> Freestream Mach number\\
$n$ \>\> Number of samples\\
$N$ \>\> Number of structural nodes\\
$p$ \>\> Polynomial order\\
$\mathcal{P}^p$ \>\> Polynomial expansion operator\\
$\mathcal{P}^p_s$ \>\> Sparse polynomial expansion operator\\
$\bm{Q}$ \>\> Generalized aerodynamic force vector\\
$q_\infty$ \>\> Dynamic pressure [Pa]\\
$s$ \>\> Number of non-zero coefficients in $\bm{d_s}$\\
$\mathcal{T}$ \>\> Multi-variable Taylor series operator\\
$\bm{v}$, $\bm{\dot{v}}$, $\bm{\ddot{v}}$ \>\> Displacement, velocity and acceleration vectors  (nodal coordinates)\\
$\alpha_0$ \>\> Freestream angle-of-attack [$^\circ$]\\
$\alpha$ \>\> Pitch angle [$^\circ$]\\
$\kappa$ \>\> Total number of coefficients in $\bm{d}$\\
$\bm{\Phi}$ \>\> Matrix of normal modes\\
$\omega$ \>\> Natural frequency\\
$\bm{\xi}$, $\bm{\dot{\xi}}$, $\bm{\ddot{\xi}}$ \>\> Generalized displacement, velocity and acceleration vectors\\

\end{tabbing}

	\section{Introduction}

        In recent decades, exponential growth in computing power and rapid advancements in the field of machine learning have enabled new technologies that are having a major impact on the aerospace industry. One such technology being broadly adopted by aircraft programs is the digital twin paradigm. Broadly speaking, an aircraft digital twin aims to digitally replicate operational conditions, including the vast array of subsystem and component level physics, via multi-physics simulation. Despite the progress, a true aircraft digital twin remains elusive. This is predominantly due to coupled high-fidelity numerical modeling of the governing physics being computationally intractable. A more feasible concept is the bottom up approach to digital twinning where, starting with a reduced order representation of the aircraft, physics can be added as needed depending on the desired analysis. In terms of aeroelasticity and airframe loads assessment, there is a trajectory towards aeroelastic flight simulator technology. A recent example is the \enquote{stick-to-stress} concept, coupling the rigid-body flight dynamics and aeroelastic equations-of-motion (EoM) in a reduced order framework to model aeroelastic stability and critical airframe stresses~\cite{lindsley12}.  

        Aeroelastic problems can often be well described by linearized aeroelastic reduced order models (ROMs). However, for some problems, such as, those concerning nonlinear transonic flow phenomena, linearized ROMs can be overly conservative and  inaccurate. An alternative is to use high-fidelity computational fluid dynamics (CFD) codes, however, this can be computationally exhaustive considering that aeroelastic aspects of aircraft design and certification require a large number of simulations. Using CFD becomes computationally intractable when considering an aeroelastic flight simulator. Nonlinear ROMs for unsteady aerodynamic and aeroelastic systems are an efficient alternative which have progressed significantly in the last half-century. 
        
        The range of nonlinear ROM methods is vast, including; polynomial functionals~\cite{silva05}, proper orthogonal decomposition~\cite{hall00}, and harmonic balance~\cite{thomas04a}. Naturally, nonlinear ROMs based on deep neural networks (DNN) have been proposed recently, an example for parametric modeling of nonlinear aeroelastic systems is provided in~\cite{li19}. However, for practical nonlinear aeroelastic applications DNN-based approaches may require excessive training data. Another recent advancement is in the use of dynamic mode decomposition (DMD) where unsteady aerodynamic pressures on a lifting surface are projected onto a lower dimensional subspace and can be used in an aeroelastic framework~\cite{fonzi24, yao24}. Although the use of DMD enables physically interpretable models with impressive accuracy, for some problems it may add unnecessary complexity and computational cost related to storing and working with distributed quantities.  
        


        Approaches based on polynomial functionals, typically a multi-variable Taylor series (or Volterra series), are particularly well suited to nonlinear transonic aeroelastic problems, having been pioneered by Silva~\cite{silva97} nearly thirty years ago. This class of ROM describes the nonlinear unsteady aerodynamic forces on a body as a function of bodies motion (rigid or elastic). The earliest works in nonlinear aeroelasticity consider the identification of Volterra kernels by perturbing the full-order model (FOM) using impulse/step functions and directly recording the nonlinear aerodynamic impulse responses (kernels)~\cite{silva99,silva04,hong03,raveh01, marzocca04, marzocca05,balajewicz10}. Although this means of identifying the kernels has rich theoretical foundations, it does come with limitations in terms of nonlinear modeling. More specifically, this includes $i$) dynamic mesh issues that arise from applying high amplitude impulses in CFD codes, and $ii$) a large computational burden associated with identifying higher-order kernels (known as the curse-of-dimensionality). This generally limited the application of the traditional Volterra series approach to the identification of second-order systems. More recent applications of Volterra theory in nonlinear aeroelasticity consider the identification of the linear and nonlinear kernels from input-output relations~\cite{ balajewicz09, balajewicz12, depaula15, depaula19,levin22, brown22, candon24a, candon24b, candon24c} allowing robust identification of higher-order systems and enabling simplistic extension to multi-input identification. However, this alone does not overcome the curse-of-dimensionality that exists when considering multi-variable polynomial fitting problems. To this effect, sparsity promotion becomes essential. Balajewicz and Dowell~\cite{balajewicz12} employ rigid sparsity, where the sparsity pattern is pre-selected and embedded prior to identifying the ROM. Levin~\textit{et al.}~\cite{levin22} propose a formulation where the linear impulse response of the system is fixed and second-order terms are added based on the rigid sparsity approach mentioned above. In a similar formulation, Brown \textit{et al.}~\cite{brown22} use $\ell_1$-regularized least-squares to identify the nonlinear terms. In recent work of the authors~\cite{candon24a, candon24b} a single-input ROM formulation is proposed where rigid sparsity is compared to $\ell_1$-regularized least-squares and orthogonal matching pursuit (OMP) for the automatic identification of an optimal sparse subset of nonlinear coefficients up to order three. The ROM based on OMP is shown to outperform the other two sparsity promoting techniques, demonstrating good performance for complex three-dimensional aeroelastic problems. The authors go on to propose a method for parameterizing the approach in~\cite{candon24c} using Lagrange interpolation. 
        
        

        This paper proposes an OMP-based formulation for the automatic identification of optimized sparse multi-input Volterra coefficients. The benefit of the multi-input formulation is two-fold, $i$) only one CFD simulation is required to identify the ROM coefficients for all modes, and $ii$) nonlinear cross-terms can be identified which, for some systems, significantly improves accuracy~\cite{balajewicz10,balajewicz12,depaula19}. The drawback of the multi-input formulation is that it is a secondary avenue for the curse of dimensionality, potentially increasing the offline computational cost exponentially with the number of structural modes being considered. 
        
        Herein, it is shown that through the proposed Optimally Sparse Multi-input ROM (OSM-ROM), drawbacks related to the curse-of-dimensionality can be avoided completely, and the multi-input formulation can be used with no added offline computational cost. Using multi-input generalized displacements to excite the full-order CFD model, the generalized aerodynamic forces are recorded as outputs and OMP is used to automatically select an small optimal subset of the tens-of-thousands of total possible ROM coefficients. This means that the only consideration in the generation of training data is to ensure that the range of amplitudes and frequencies that will be observed in the use cases are adequately represented, while the number of training samples is implicitly accounted for. The case study considers the Benchmark Supercritical Wing (BSCW)~\cite{dansberry92, dansberry93, piatek03}, using cases from the Aeroelastic Prediction Workshops (AePW)~\cite{chwalowski17, chwalowski24}. The results demonstrate that the OSM-ROM performance is excellent relative to the FOM including for forced response, flutter and supercritical limit cycle oscillations (LCO). The robustness and limitations of the proposed method are assessed using cases of increasing complexity, including varying degrees of shock wave motion and boundary layer separation, achieved by increased the Mach number and angle-of-attack (AoA).

    \section{Formulation of the Optimized Sparse Multi-Input Polynomial Functionals}
    \label{sec:roms}

    In this section, the procedure for identifying the unsteady aerodynamic ROM is described. This definition is for multi-input single-output (MISO) identification, $i.e.$, terms describing nonlinear interactions between structural modes will be identified if nonlinear cross-coupling exists. Alternatively, if nonlinear cross-coupling effects are not present, the OMP-based formulation should not identify cross-terms.  

    \subsection{Polynomial Functional Approximation of the Unsteady Aerodynamic Forces}
     \label{sec:tsexp}
 
    For a nonlinear time-invariant (NLTI) system, a reasonable assumption is that the unsteady aerodynamic forces on a structure can be described in discrete time as a dynamic function of the displacements of a subset of $m$ normal modes by

    \begin{equation}
    \label{eq:TS2}
        \bm{Q}[n] = f(\bm{\xi})
    \end{equation}

    \noindent where

    \begin{equation}
        \bm{\xi} = \{\bm{\xi}_1, \hdots, \bm{\xi}_m\}^T
    \end{equation}

    \noindent and $\bm{\xi}_m = \{\bm{\xi}_m[n-1], \bm{\xi}_m[n-2], \hdots \bm{\xi}_m[n-k]\}$ is a vector of generalized displacements of mode $m$, $k$ defines the number of time lags, $\bm{Q}\{n\}$ represents the generalized aerodynamic force vector at time interval $n$, and $f(\bm{\xi})$ is an unknown nonlinear dynamic function. Then, provided that the system is mildly nonlinear and memory fading $f(\bm{\xi})$ can be approximated by the multi-variable Taylor series expansion at a reference location $\bm{\xi} = \bm{a}$, given in multi-index form as~\cite{duistermaat10}
   

    \begin{equation}
        \label{eq:TS5}
            f(\bm{\xi}) \approx \mathcal{T}(\bm{\xi}) = f(\bm{a}) + \sum_{p_i=1}^p \frac{1}{{p_i}!}(\bm{\xi} - \bm{a})^{p_i}\bm{D}^{p_i}
     \end{equation}

    \noindent where $\bm{D}^{p_i} = (\partial^{p_i} f)(\bm{a})$ is a tensor of partial derivatives of $f(\bm{a})$ of order $p_i$. Specifically, $\bm{D}^1$ is the gradient of $f(\bm{a})$, $\bm{D}^2$ is the Hessian matrix, and so on. Given that $f(\bm{a})$ is not known a priori, the coefficients of $\bm{D}^{p_i}$ are estimated from input-output training data, $i.e.$, the coefficients are identified to minimize the error between the Taylor approximation and the true values by $min||\bm{Q}\{n\} - \mathcal{T}(\bm{\xi})||$. The remainder of this section is dedicated to describing the procedure for identifying the coefficients. 

    \subsection{Multi-Input and Output Matrices}

    \subsubsection{Band Limited Random Excitation}
    
    The first step in creating the unsteady aerodynamic ROM is to perturb the structural model via the full-order aerodynamic solver and to record the aerodynamic responses. Using the MISO identification procedure, each mode is excited at the same time using uncorrelated band limited random noise. 
    
    An important consideration is to observe a smooth transition from the undeformed structure and converged steady-state fluid forces. The reason being that any discontinuity ($i.e.$, a step-like change in structural displacement) will cause spurious aerodynamic response information and inaccuracies in the ROM identification process. A hyperbolic tangent is applied to the the raw signal $\bm{\xi}_{j,r}$, ensuring a smooth transition to the modal excitation over the first 100 time intervals, given for $n$ total input samples by

        \begin{equation}
        \label{eq:smramp}
            \bm{\xi}_j = \left(0.5\tanh{\left( \frac{n-100}{20} \right)}+1\right)\bm{\xi}_{j,r}
        \end{equation}

    \noindent where $\bm{\xi}_j$ is the band limited random signal used to excite the $j^{th}$ mode. 

    \subsubsection{Nonlinear Multi-Input Matrix}

     Considering a total of $m$ modes as inputs, lower left triangular circulant matrices are constructed for the input to each $j^{th}$ mode using $\bm{\xi}_j$ (truncated for $k$ time lags) and concatenated, giving:

    \begin{equation}
             \bm{L} = [\bm{L}^{1}, \bm{L}^{2} \cdots \bm{L}^{m}]\in \mathbb{R}^{n \times mk}
     \end{equation}
     
     \noindent where
     
     \begin{align} 
        \label{eq:TS8}
        \bm{L}^j = \begin{bmatrix}
        \bm{\xi}_{j}[0] & 0 & 0 & \hdots & 0\\
        \bm{\xi}_{j}[1]  & \bm{\xi}_{j}[0] & 0 & \hdots &0 \\
        \vdots  &\vdots & \vdots& \ddots & \vdots \\
        \bm{\xi}_{j}[n-1]  & \bm{\xi}_{j}[n-2]  & \bm{\xi}_{j}\{n-3]  &\hdots & \bm{\xi}_{j}[n-k]\\
      \end{bmatrix}\in \mathbb{R}^{n \times k}, j \in \mathbb{N}[1,m]
      \end{align}

     \noindent and the nonlinear multi-input matrix is obtained by computing the $p^{th}$-order monomials of the rows:

     \begin{align}
             \bm{\mathcal{M}} & = \mathcal{P}^p(\bm{L}_{l}), l\in\mathbb{N}[1,n] \\
             & = [\bm{\mathcal{M}}_1^{j_1} \cdots \bm{\mathcal{M}}_2^{j_1j_2} \cdots \bm{\mathcal{M}}_p^{j_1 \hdots j_p}]\in \mathbb{R}^{n \times \kappa}, \ j_1 \hdots j_p \in \mathbb{N}[1,m]
     \end{align}

     \noindent where $\mathcal{P}^p()$ denotes the operation to calculate the $p^{th}$-order monomials of $\bm{L}$, $\kappa = \sum_{p_i=1}^{p} \binom{mk+(p_i-1)}{p_i}$, and $\bm{\mathcal{M}}$ contains all monomials of the inputs in $\bm{L}$ up to order $p$. The first-order matrices $\bm{\mathcal{M}}^{j_1}_1 \equiv \bm{L}^j$. Of the nonlinear matrices ($p>1$), when $j_1 = j_2 = \hdots j_p$ the matrix $\bm{\mathcal{M}}_p^{j_1 \hdots j_p}$ contains direct nonlinear monomials of $\bm{L}^j$. On the other hand, when the superscripts are not equal the matrix $\bm{\mathcal{M}}_p^{j_1 \hdots j_p}$ contains the cross nonlinear monomials of $\bm{L}^{j_1} \hdots \bm{L}^{j_p}, \ j_1\hdots j_p\in\mathbb{N}[1,m]\ | \ \neg \ (j_1 = j_2 =  \hdots j_p)$. The nonlinear multi-input input matrices are defined explicitly as~\cite{balajewicz12}

     \begin{equation}
     \begin{split} 
        \label{eq:TS8}
        \bm{\mathcal{M}}_2^{j_1 j_2} =   \begin{bmatrix}
        \bm{\xi}_{j_1}[0] \cdot \bm{\xi}_{j_2}[0] & 0 & 0 & \hdots \\
        \bm{\xi}_{j_1}[1] \cdot \bm{\xi}_{j_2}[1] & \bm{\xi}_{j_1}[1] \cdot \bm{\xi}_{j_2}[0]& \bm{\xi}_{j_1}[0] \cdot \bm{\xi}_{j_2}[0] & \hdots \\
        \vdots  &\vdots & \vdots&   \\
        \bm{\xi}_{j_1}[n-1] \cdot \bm{\xi}_{j_2}[n-1]  & \bm{\xi}_{j_p}[n-2]\cdot \bm{\xi}_{j_p}[n-1]  & \bm{\xi}_{j_p}[n-2]\cdot \bm{\xi}_{j_p}[n-2]  &\hdots \\
      \end{bmatrix}
      \end{split}
      \end{equation}
      
    \begin{equation}
     \begin{split} 
        \label{eq:TS8}
        \bm{\mathcal{M}}_p^{j_1 \hdots j_p} =   \begin{bmatrix}
        \prod_{n=1}^{p}\{\bm{\xi}_{j_n}[0]\} & 0 & 0 & \hdots \\
        \prod_{n=1}^{p}\{\bm{\xi}_{j_n}[1]\}  & \prod_{n=1}^{p-1}\{\bm{\xi}_{j_n}[1]\}\bm{\xi}_{j_p}[0] & \prod_{n=1}^{p-2}\{\bm{\xi}_{j_n}[1]\}\bm{\xi}_{j_p}[1] & \hdots \\
        \vdots  &\vdots & \vdots&   \\
        \prod_{n=1}^{p}\{\bm{\xi}_{j_n}[n-1]\}  & \prod_{n=1}^{p-1}\{\bm{\xi}_{j_p}[n-2]\}\bm{\xi}_{j_p}[n-1]  & \prod_{n=1}^{p-1}\{\bm{\xi}_{j_p}[n-2]\}\bm{\xi}_{j_p}[n-2]  &\hdots \\
      \end{bmatrix}
      \end{split}
      \end{equation}


     \subsubsection{Output Matrix}

     Using the full-order aerodynamic solver, the vector of outputs is computed by exciting each $j^{th}$ mode at the same time with the uncorrelated inputs $\bm{\xi}_j \in \mathbb{R}^n, j \in \mathbb{N}[1,m]$ and the full-order aerodynamic forces are projected onto each $i^{th}$ mode $\bm{Q}^i(\bm{\xi}_1, \hdots, \bm{\xi}_m) \in \mathbb{R}^{n}, i \in \mathbb{N}[1,m]$, giving the matrix of outputs:
     
     \begin{equation} 
        \bm{Q} = [\bm{Q}^1(\bm{\xi}_1, \hdots,\bm{\xi}_m) , \bm{Q}^2(\bm{\xi}_2, \hdots,\bm{\xi}_m), \hdots, \bm{Q}^m(\bm{\xi}_1, \hdots,\bm{\xi}_m)]\in \mathbb{R}^{n \times m}
      \end{equation}

    \subsection{ROM Identification}

    Identifying the coefficients of the ROM is a linear problem~\cite{rugh81} which can be defined for the nonlinear multi-input matrix $\bm{\mathcal{M}}$ and corresponding aerodynamic force output matrix $\bm{Q}$ by
    
    \begin{equation}
    \label{eq:linprob}
        \bm{\mathcal{M} d} = \bm{Q} 
    \end{equation}
    
    \noindent where $\bm{d} = [\bm{d}^1,\bm{d}^2, \hdots, \bm{d}^m ] \in \mathbb{R}^{\kappa \times m}$ contains the partial derivatives (linear, direct nonlinear and nonlinear cross-terms) which are unknown. Although any standard least squares (LS) approach can be used to solve the inverse problem for $\bm{d}$, the number of terms in $\bm{d}$ to be identified grows exponentially with the order of the ROM, the number of time lags and the number of structural modes. Considering that the required number of samples in $\bm{Q}$ is proportional to the number of terms to be identified in $\bm{d}$ this can quickly become computationally intractable given that training data is generated using a CFD code. Thus, from a practical perspective for 3D aeroelastic problems, sparsity promotion is essential. 
    

    
    \subsubsection{Identification using Least Squares}

    The dense set of terms (no sparsity) of the \textbf{M}ulti-input \textbf{ROM} (M-ROM) are identified using LS, evaluated according to
     
    \begin{equation}
        \bm{d} = \bm{\mathcal{M}{^+}Q} \in  \mathbb{R}^{\kappa \times m}
    \end{equation}

    \noindent where $^+$ is the Moore-Penrose Pseudo-Inverse and the partial derivatives for the $i^{th}$ mode are given by:

    \begin{equation}
    \bm{d}^{i} = \{\bm{d}_1^{i,j_1} \cdots \bm{d}_2^{i,j_1j_2} \cdots \bm{d}_p^{i,j_1\hdots j_p}\}^T \in \mathbb{R}^{\kappa}, \ i \in \mathbb{N}[1,m], \ j_1 \hdots j_p \in \mathbb{N}[1,m]
    \end{equation}
    
    \noindent which contains the linear and nonlinear partial derivatives corresponding to the generalized aerodynamic forces for $\bm{Q^i}(\bm{\xi_1}, \hdots,\bm{\xi_m}) \in \mathbb{R}^{n}, i \in \mathbb{N}[1,m]$. Of the nonlinear terms ($p>1$), when $j_1 = j_2 = ... =j_n$, $\bm{d}_p^{i,j_1\hdots j_p}$ contains direct nonlinear terms which do not describe nonlinear interactions between modes. Conversely, when $\neg j_1 = j_2 = ... =j_n$, $\bm{d}_p^{i,j_1\hdots j_p}$ contains the nonlinear cross-terms. This approach is only used to identify the first-order ROMs in this paper. 
    
    \subsubsection{Identification using Orthogonal Matching Pursuit}
    \label{sec:OSROM}
    
    OMP is a greedy algorithm that recovers a sparse representation of a signal in a step-by-step iterative manner~\cite{foucart13}. For an extended definition of the OMP algorithm used in this work, the reader is referred to recent work by the authors~\cite{candon24a}. The objective of OMP is to identify a sparse representation of $\bm{d}$, denoted by $\bm{d_s}$, by solving the NP-hard problem:

    \begin{equation*}
    \label{eq:omp}
        \underset{\bm{{d_s}}}{\operatorname{argmin}}  ||\bm{{d_s}}||_0 \quad \textrm{subject to} \quad \bm{\mathcal{M}d_s} = \bm{Q}
    \end{equation*}

    \noindent where $||\bm{d_s}||_0$ is the $\ell_0$ pseudo-norm, or the number of non-zero elements in $\bm{d_s}$. Assuming that $\bm{d_s}$ is $s$-sparse ($s_{d_s} \geq ||\bm{d_s}||_0$), it can be recovered exactly by OMP if $\bm{\mathcal{M}}$ and $\bm{d_s}$ satisfy following inequality:

    \begin{equation}
    \label{eq:TS17}
    \mu_\mathcal{M} < \frac{1}{2s_{d_s} - 1}
    \end{equation}

     \noindent where $\mu_\mathcal{M}$ is the mutual coherence of the columns of $\bm{\mathcal{M}}$ and $s_{d_s}$ is the sparsity of $\bm{d_s}$. From Eq.~\ref{eq:TS17}, $\bm{d_s}$ can be at most $\frac{1}{2\mu_\mathcal{M}}$-sparse. Using the number of non-zero terms as the stopping criterion, the NP-hard problem can be solved directly using the standard OMP routine available in the Python sklearn library, giving the \textbf{O}ptimal \textbf{S}parsity \textbf{M}ulti-input \textbf{ROM} (OSM-ROM):

    \begin{equation} 
        \bm{d_s} = \text{OMP}(\bm{\mathcal{M}}, \bm{Q}) =  \ [\bm{d_s^1}, \bm{d_s^2}, \hdots, \bm{d_s^m}] \in \mathbb{R}^{\kappa \times m}
      \end{equation}
    
    \noindent where 

    \begin{equation}
    \bm{d_s}^{i} = \{ {\bm{d_s}}_1^{i,j_1} \cdots {\bm{d_s}}_2^{i,j_1j_2} \cdots {\bm{d_s}}_p^{i,j_1\hdots j_p}\}^T \in \mathbb{R}^{\kappa}, \ i \in \mathbb{N}[1,m], \ j_1 \hdots j_p \in \mathbb{N}[1,m]
    \end{equation}

    \noindent which contains a sparse representation of the linear and nonlinear partial derivatives ($s<<\kappa)$, corresponding to the generalized aerodynamic forces for $\bm{Q^i}(\bm{\xi_1}, \hdots,\bm{\xi_m}) \in \mathbb{R}^{n}, i \in \mathbb{N}[1,m]$. Of the nonlinear terms ($p>1$), when $j_1 = j_2 = ... j_p$, $ {\bm{d_s}}_{p}^{i,j_1\hdots j_p}$ contains direct nonlinear terms which do not describe nonlinear interactions between modes. Conversely, when $\neg j_1 = j_2 = ... j_p$, ${\bm{d_s}}_p^{i,j_1\hdots j_p}$ contains the nonlinear cross-terms. This approach is used to identify the nonlinear ROMs in this paper.


    



	\section{Nonlinear Aeroelastic Framework}
	
	\subsection{Benchmark Supercritical Wing}
	\label{sec:model} 
	
\begin{figure}[h]
		\centering
		\subfigure[]{\label{bscw_oml}
			\includegraphics[width=0.4\textwidth]{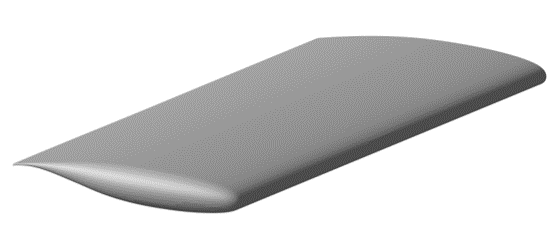}}
		\subfigure[]{\label{bscw_str}
			\includegraphics[width=0.4\textwidth]{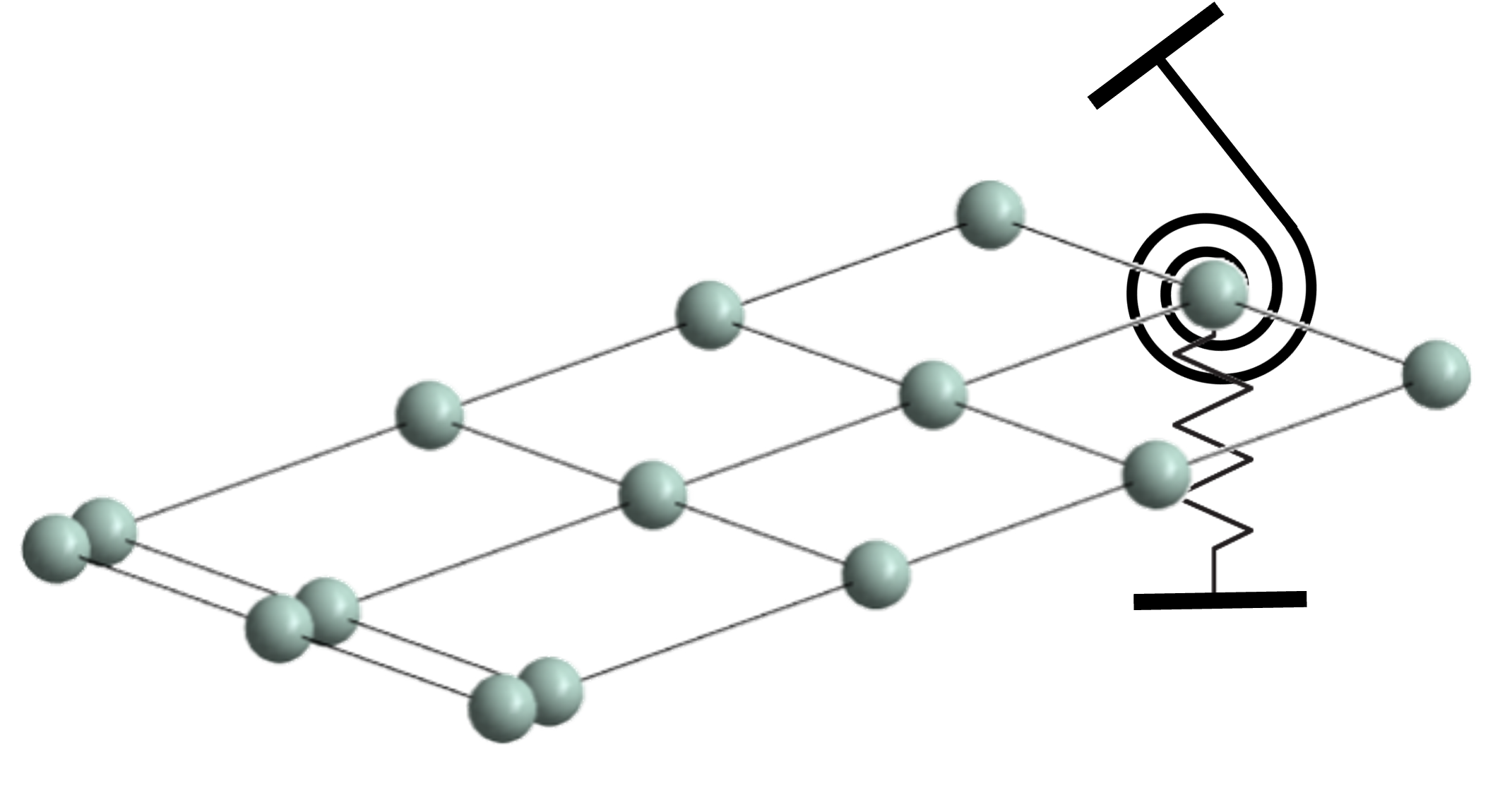}}
		\caption{BSCW model: a) fluid outer mould line and b) structural model}
		\label{fig:1}
	\end{figure}

        The benchmark supercritical wing (BSCW) is a benchmark aeroelastic model with experiments conducted at the NASA Transonic Dynamics Tunnel, used as a test subject in the AIAA Aeroelastic Prediction Workshops (AEPW). The model consists of a rectangular wing with a NASA SC(2)-0414 airfoil section and a wing tip cap, shaped as a tip of revolution, as is presented in Fig~\ref{bscw_oml}. Comprehensive experimental data is available at various transonic operating conditions for the stationary wing, forced pitch oscillations performed on an oscillating turntable (OTT)~\cite{piatek03}, and for flutter where the wing is mounted on a pitch and plunge apparatus (PAPA)~\cite{dansberry92}. The wing is considered to be rigid, with longitudinal and rotational springs attached to the root center node, which is free to move in the heave and pitch directions, while constrained in all other degrees-of-freedom (Fig.~\ref{bscw_str}). The BSCW model parameters, structural and fluid models, and experimental data are available on the AEPW website.

	\subsection{Nonlinear Aeroelastic Equation-of-Motion}
	The undamped equation-of-motion for an aeroelastic system in nodal coordinates is given as
	\begin{equation}
		\bm{M}\ddot{\bm{v}} + \bm{K}(\bm{v}) + \bm{F} = 0 \label{eq1}
	\end{equation}

	\noindent where $\bm{M}$ and $\bm{K}$ are the structural mass and stiffness matrices, $\bm{v} = \{v_1,\ v_2,\ \ldots,\ v_N\}^T$ is the displacement vector of $N$ degrees-of-freedom, and $\ddot{\bm{v}}$ is the second time derivative of $\bm{v}$. The external force vector $\bm{F} = \{F_{1},\ F_{2},\ \ldots,\ F_{N}\}^T$ describes the aerodynamic force vector in nodal coordinates. The system described by Eq.~\eqref{eq1} can be reduced to generalized coordinates, such that the structural motion is approximated as the linear superposition of a subset of $m$ normal modes $\bm{\Phi}$ due to generalized displacement $\bm{\xi}$, given by:
    
    \begin{equation}
    \label{eq:modalAE}
    \bm{GM}\bm{\ddot{\xi}} + \bm{GK}(\bm{\xi}) + \bm{Q} = 0 
    \end{equation}

    \noindent where $\bm{GM} = \bm{\Phi}^T\bm{M_v}\bm{\Phi}$ and $\bm{GK} = \bm{\Phi}^T\bm{K}\bm{\Phi}$ are the generalized mass and stiffness matrices respectively, and and $\bm{Q} = \bm{\Phi}^T\bm{F}$ is the generalized aerodynamic force vector. The heave and pitch modes for the BSCW are presented in Figs.~\ref{mode1_free} and~\ref{mode2_free}.


	\subsection{Computational Fluid Dynamics Model}
\begin{figure}[h]
		\centering
		\subfigure[heave mode (3.33 Hz)]{\label{mode1_free}
			\includegraphics[width=0.32\textwidth]{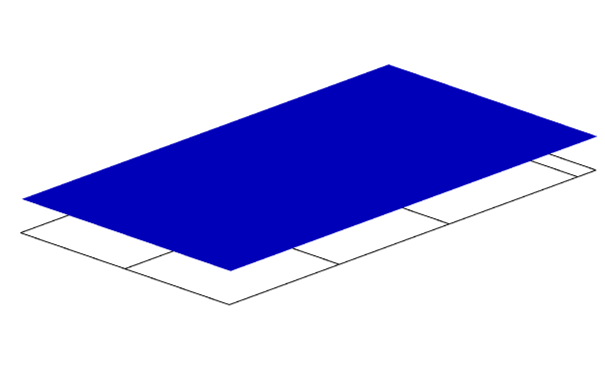}}
		\subfigure[pitch mode (5.2Hz)]{\label{mode2_free}
			\includegraphics[width=0.32\textwidth]{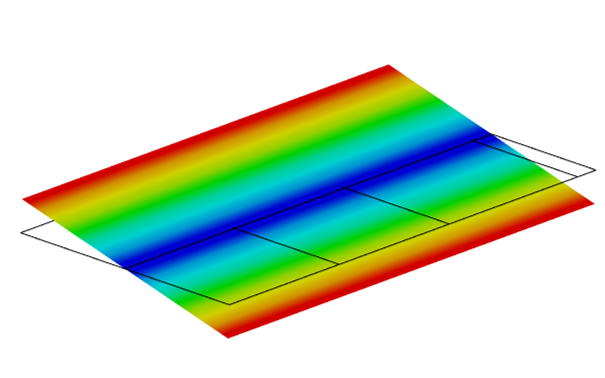}}
        \subfigure[structured fluid mesh]{\label{cfd_mesh}
            \includegraphics[width=0.32\textwidth]{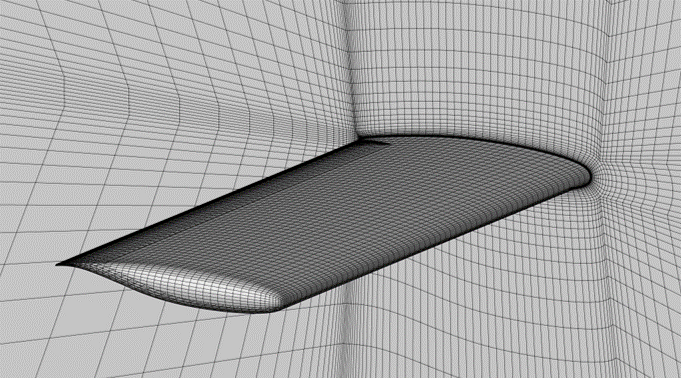}}
		\caption{BSCW structural modes and fluid mesh}
		\label{fig:3}
	\end{figure}

	For full-order simulations the generalized aerodynamic forces $\bm{Q}$ are obtained using the commercial finite-volume Navier-Stokes solver ANSYS Fluent 2023 R1. The RANS equations for transient flowfields are solved via a coupled pressure-based solver with implicit second-order spatial discretization for all flow variables, second-order temporal discretization, and Rhie-Chow: distance-based flux interpolation. Menter's Shear Stress Transport model (SST-$k\omega$) is used for turbulence modeling. The convergence criteria are set to $1\times10^{-5}$ for the scaled residuals at each time-step. The investigation is conducted on a structured grid of 1.48M elements, with a minimum orthogonal quality of 0.0028, as is presented in Fig.~\ref{cfd_mesh}. A transient time-step of $\Delta t = 0.001$s is used in all simulations.


    \subsection{Nonlinear Aerodynamic Reduced Order Model}

\begin{figure}[h]
        \centering
        \includegraphics[width=1\textwidth]{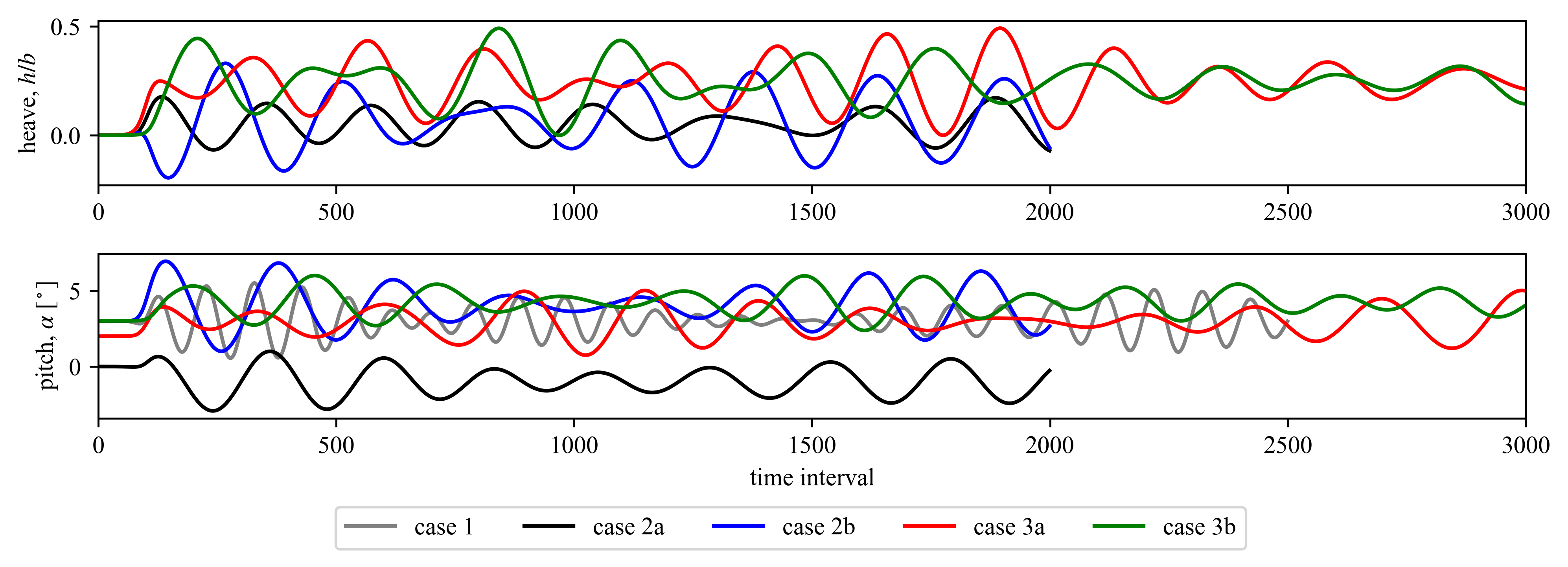}
        \caption{Band limited random excitation used for each case}
        \label{fig:aeroInp}
    \end{figure}

    In this paper ROMs are generated at five separate operating conditions. The ROM parameters, use cases and band limited excitation statistics for each case are described in Table~\ref{tab:case_summary}, and the band limited random signal for each mode is presented in Fig.~\ref{fig:aeroInp}. For all cases the first half of the samples are used for model training and the remainder for cross-validation and hyperparameter tuning. Hyperparameter tuning is conducted using a grid search, with the objective of minimizing the error between the unsteady aerodynamic ROM and FOM solutions for the cross-validation data set. The hyperparameters considered are the number of time lags $k$ and the number of sparse coefficients to be identified by OMP, $s$. A detailed description of the hyperparameter tuning strategy is provided in recent work of the authors~\cite{candon24b}.

\begin{figure}[h]
    \centering
    \subfigure[time domain]{\label{mode1_free}
        \includegraphics[width=1\textwidth]{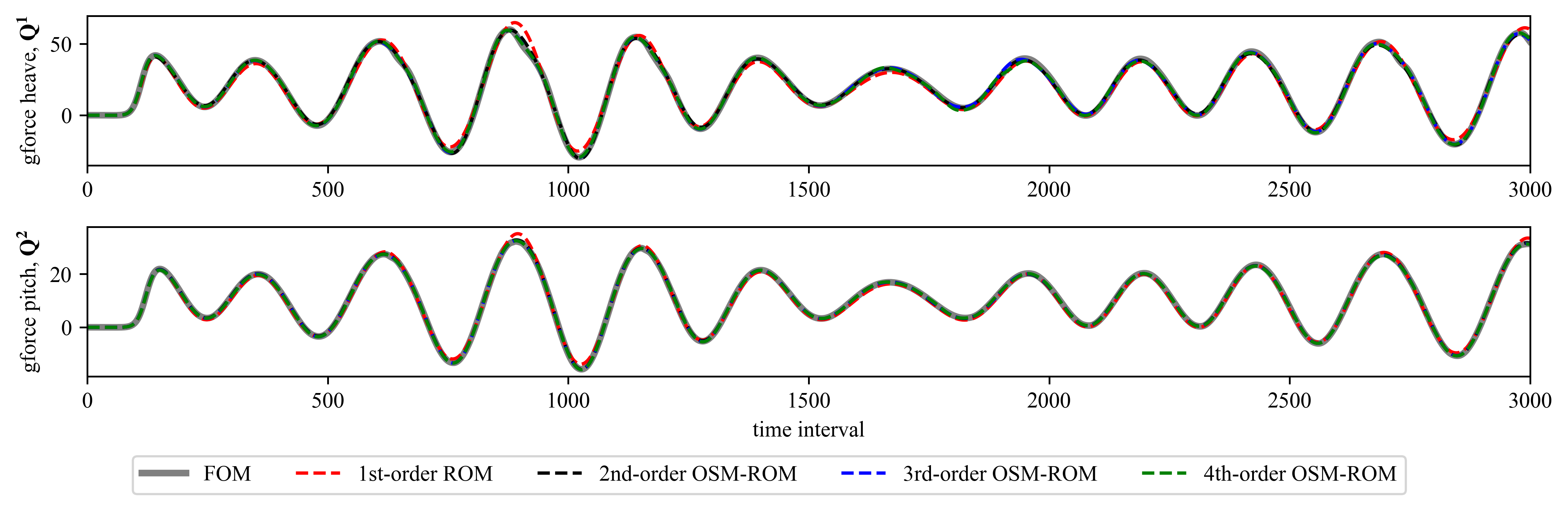}}
        
    \subfigure[frequency domain]{\label{mode2_free}
        \includegraphics[width=1\textwidth]{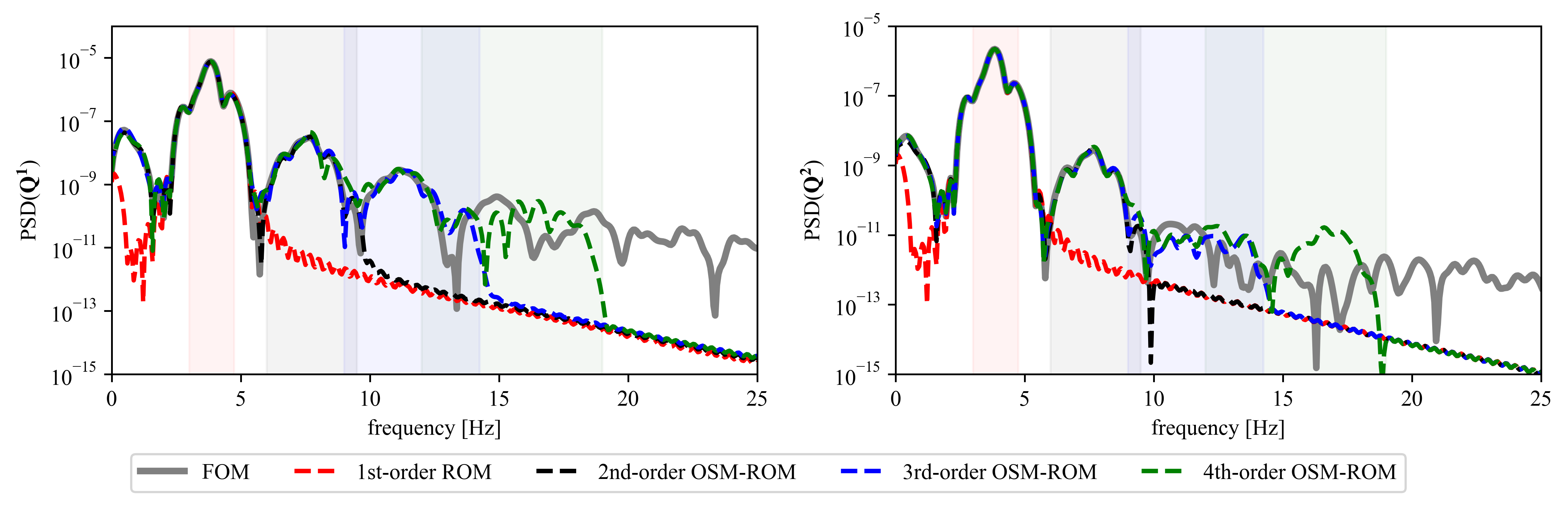}}
    \caption{Generalized force response to training data for case 3a: $M_\infty = 0.8, \ \alpha_0 = 2^\circ$}
    \label{fig:traindata_3a}
    \end{figure}

\begin{figure}[h]
    \centering
    \subfigure[time domain]{\label{mode1_free}
        \includegraphics[width=1\textwidth]{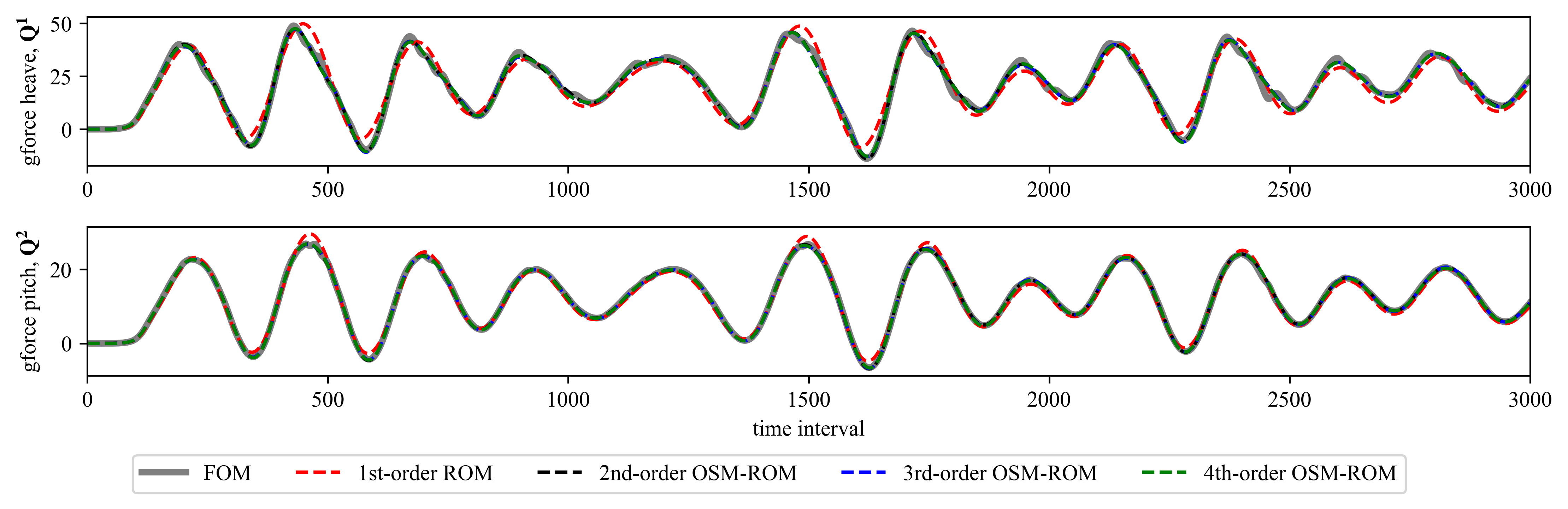}}
        
    \subfigure[frequency domain]{\label{mode2_free}
        \includegraphics[width=1\textwidth]{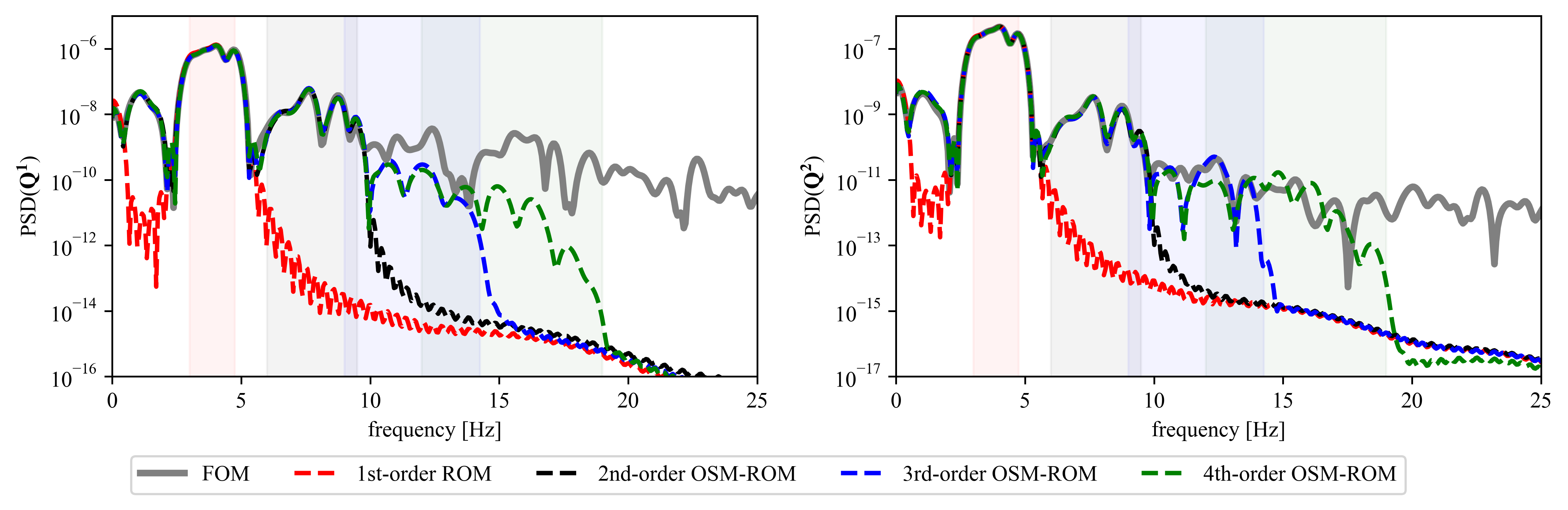}}
    \caption{Generalized force response to training data for case 3b: $M_\infty = 0.8, \ \alpha_0 = 3^\circ$}
    \label{fig:traindata_3b}
    \end{figure}

    Figure~\ref{fig:traindata_3a} presents the generalized force response to the training input signal computed the FOM and the OSM-ROMs of increasing order for case 3a. In the time-domain some discrepancy can be observed for the linear model, while all nonlinear models appear to perform equally well. Observation of the frequency domain response is more informative where it can be seen that the force response contains sub- and superharmonics of the excitation frequency band, as expected if the unsteady aerodynamic forces are nonlinear. The ROM frequency domain responses demonstrate that the 1st-order ROM can only capture the dominant frequency band, the second-order OSM-ROM also captures the second-harmonics, and so on. All nonlinear OSM-ROMs capture the subharmonics well. In heave the third harmonics are also predicted well while the performance degrades for the prediction of the fourth harmonics. In pitch, there is degradation of the prediction of both third and fourth harmonics. Figure~\ref{fig:traindata_3b} presents the generalized force responses for case 3b. This case is characterized by more unsteadiness in the training data as a result of boundary layer separation which is clearly evident in the time response of the heave mode. For this case the discrepancy between the first-order ROM and the FOM is more evident in the time response, however, again it is difficult to differentiate between the time responses of the nonlinear OSM-ROMs, noting that they appear to smooth through the unsteadiness in the forces. In the frequency domain the trends are similar to those described for case 3a with the third and fourth harmonics are captured well in the heave mode and less so in the pitch mode where they are under-predicted. 

    With almost identical time-domain generalized force responses for all nonlinear OSM-ROMs, the necessity of identifying models above order two is brought into question. An important contribution of this paper that is to understand how influential the higher harmonics are in an aeroelastic setting and therefore how important it is to identify higher-order terms.

    \begin{table}[h]
    \centering
    \begin{tabular}{ccccccccc}
        \hline
        &\multicolumn{3}{c}{operating conditions} &\multicolumn{4}{c}{training input statistics}&\\
        \hline
          case & $M_\infty$ & $\alpha_0$ [$^\circ$] & $q_\infty$ [psf] & order & $f$ [Hz] & $|h|$ / $\bar{h}$ [m] & $|\alpha|$ / $\bar{\alpha}$ [$^\circ$] & use case   \\
        \hline
        \hline
        1 & 0.7 & 3 & 170 & 1-5  & 9-11 & - & 2.5 / 0 & forced response \\
        2a & 0.74 & 0 & 162 & 1-3 & 3-5 & 0.03 / 0.01 & 2 / -1 &flutter \\
        2b & 0.74 & 3 & 169 & 1-3 & 3-5 & 0.055 / 0.025 & 3 / 1 & flutter,LCO \\
        3a & 0.8 & 2 & 180 & 1-4 & 3-5 & 0.05 / 0.03 & 2.5 / 0.5 & flutter,LCO \\
        3b & 0.8 & 3 & 218 & 1-4 & 3-5 & 0.05 / 0.05 & 2 / 1 & flutter \\
        \hline
    \end{tabular}
    \caption{Summary of ROM operating conditions, input signal statistics and use cases}
    \label{tab:case_summary}
    \end{table}
    
        \clearpage

     \subsection{Aeroelastic Time Integration}
     
    The aeroelastic system is solved using the RMIT in-house Fluid-Structure Interaction code PyFSI. Aeroelastic solutions are achieved by marching Eq.~\ref{eq:modalAE} forward in time, where the wing transient structural motion is solved using Newmark-$\beta$ time-integration. For the FOM solutions $\bm{Q}$ (Eq.~\ref{eq:modalAE}) is computed using the CFD model described above, solving for the nonlinear fluid loads at every time-step. For the aeroelastic ROM solutions, the same term is computed by solving the forward linear problem described in Eq.~\ref{eq:linprob} at each time-step. The reduced order aeroelastic EoM for a given Mach number and angle-of-attack becomes:

    \begin{equation}
    \label{eq:modalAE_rom}
    \bm{GM}\bm{\ddot{\xi}} + \bm{GK}(\bm{\xi}) + q_\infty(\mathcal{H}_0 + \bm{\mathcal{M}_{\xi}d}) = 0
    \end{equation}

    \noindent where $\bm{\mathcal{M}_{\xi}}$ contains the history of generalized displacements which is updated at every time-step. For nonlinear multi-input problems, computing the new row of $\bm{\mathcal{M}_{\xi}}$ and product $\bm{\mathcal{M}_{\xi}d}$ becomes exponentially more computationally demanding as the time integration progresses. A further benefit of the sparse multi-input ROM is that the full expansion does not need to be computed given that most of the products in $\bm{\mathcal{M}_{\xi}d_s}$ are zero. A priori knowledge of the coordinates of the non-zero terms means that a sparse representation of the multi-input matrix can be used. To avoid having to solve for the redundant combinations in $\bm{\mathcal{M}_{\xi}}$, a polynomial transformation matrix is derived in the sparse coordinate system, given by:

    \begin{equation}
        \mathcal{P}^p_s() = \mathcal{P}^{p}() | \bm{d_s} \neq 0 
    \end{equation}
    \noindent which computes only the polynomial combinations corresponding to the locations of the non-zero terms in $\bm{d_s}$. At each time-step the new row of the sparse multi-input matrix is computed according to:

    \begin{equation}
        \label{eq:modalAE_rom}
        {\bm{\mathcal{M}_{\xi s}}}_{l} = \mathcal{P}^p_s(\bm{L_{l}}) \in \mathbb{R}^{s}, l = 1, \hdots, n_{tot}
    \end{equation}

    \noindent where $n_{TOT}$ is the total number of time intervals in the simulation and $l$ represents the current time interval, meaning that only $s = ||\bm{d_s}||_0$ polynomial combinations are computed at every time interval, which is orders of magnitude smaller than $\kappa = \sum_{p_i=1}^{p} \binom{mk+(p_i-1)}{p_i}$ in the dense coordinate system. The reduced order aeroelastic EoM for a given Mach number and angle-of-attack in the sparse coordinate system becomes:

    \begin{equation}
    \label{eq:modalAE_rom}
    \bm{GM}\bm{\ddot{\xi}} + \bm{GK}(\bm{\xi}) + q_\infty(\mathcal{H}_0 + \bm{\mathcal{M}_{\xi s}d_{ss}}) = 0
    \end{equation}

    \noindent where $\bm{d_{ss}} = \bm{d_{s}} \in \mathbb{R} | \bm{d_{s}}\neq 0$ is the sparse ROM in the sparse coordinate system and only $s << \kappa$ products are computed at each time-step.

    \section{Results and Discussion}
    \label{sec:res}
    
    In this section the results for the cases described in Table~\ref{tab:case_summary} are presented and discussed, and the computational performance of the nonlinear multi-input ROM is assessed. The first-order M-ROM refers to the first-order approximation to the nonlinear system identified using LS and the $p^{th}$-order OSM-ROM refers to the nonlinear multi-input ROMs with optimal sparsity. It should be noted that through the AEPW 2 and 3, it has been shown that time-accurate flutter predictions of the BSCW are sensitive to time-step and mesh density. Time-step and mesh convergence studies have not been conducted in this work. Nevertheless, in this paper, the full-order aeroelastic model (FOM) is considered the truth model and the objective is to match this with the various ROMs - a valid approach given the nature of this work.

    \subsection{Validation}

\begin{figure}[h]
        \centering
        \subfigure[$M_\infty = 0.7$]{\label{valid1}
            \includegraphics[width=0.47\textwidth]{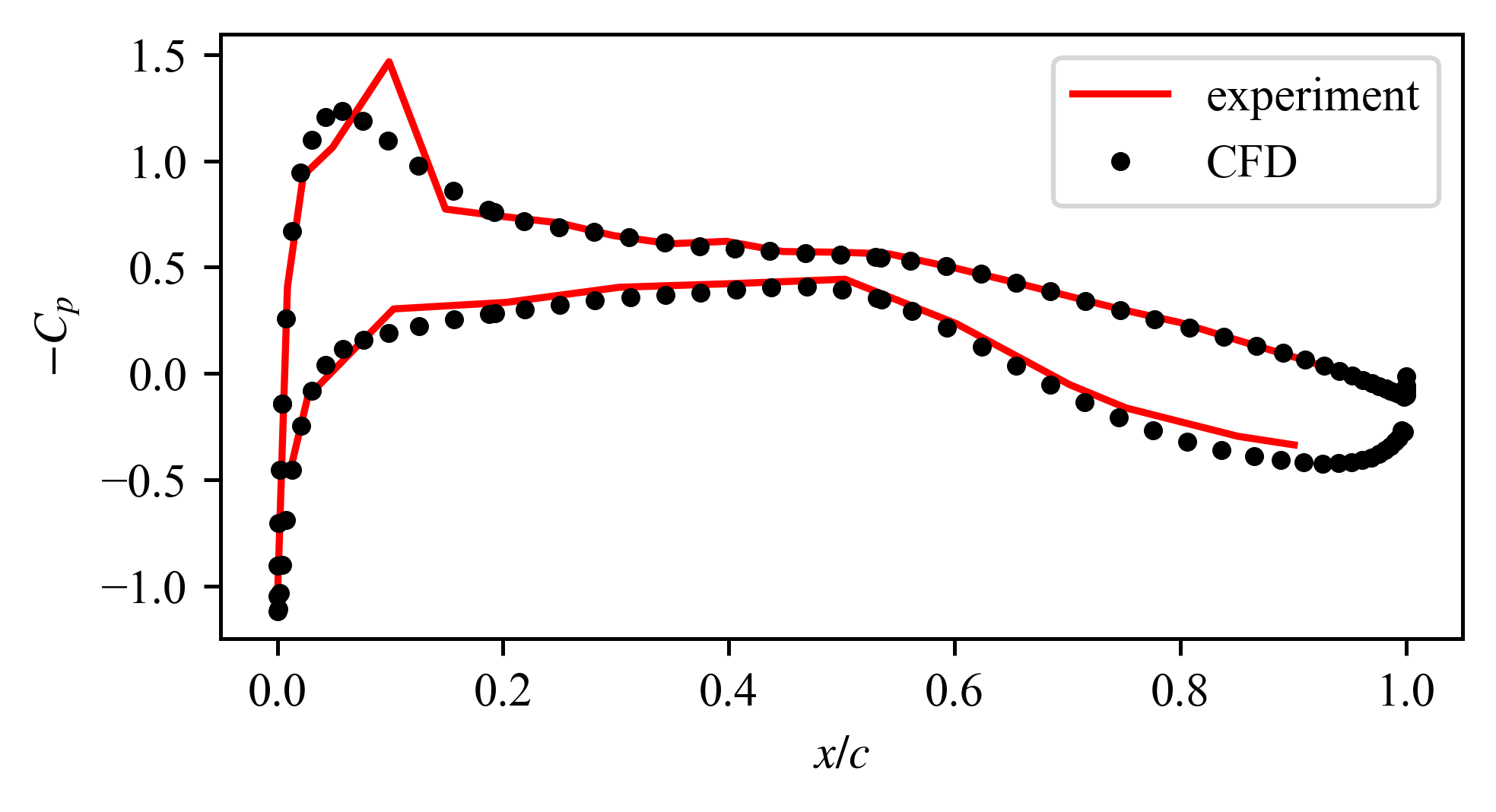}}
            \subfigure[$M_\infty = 0.74$]{\label{valid2}
            \includegraphics[width=0.47\textwidth]{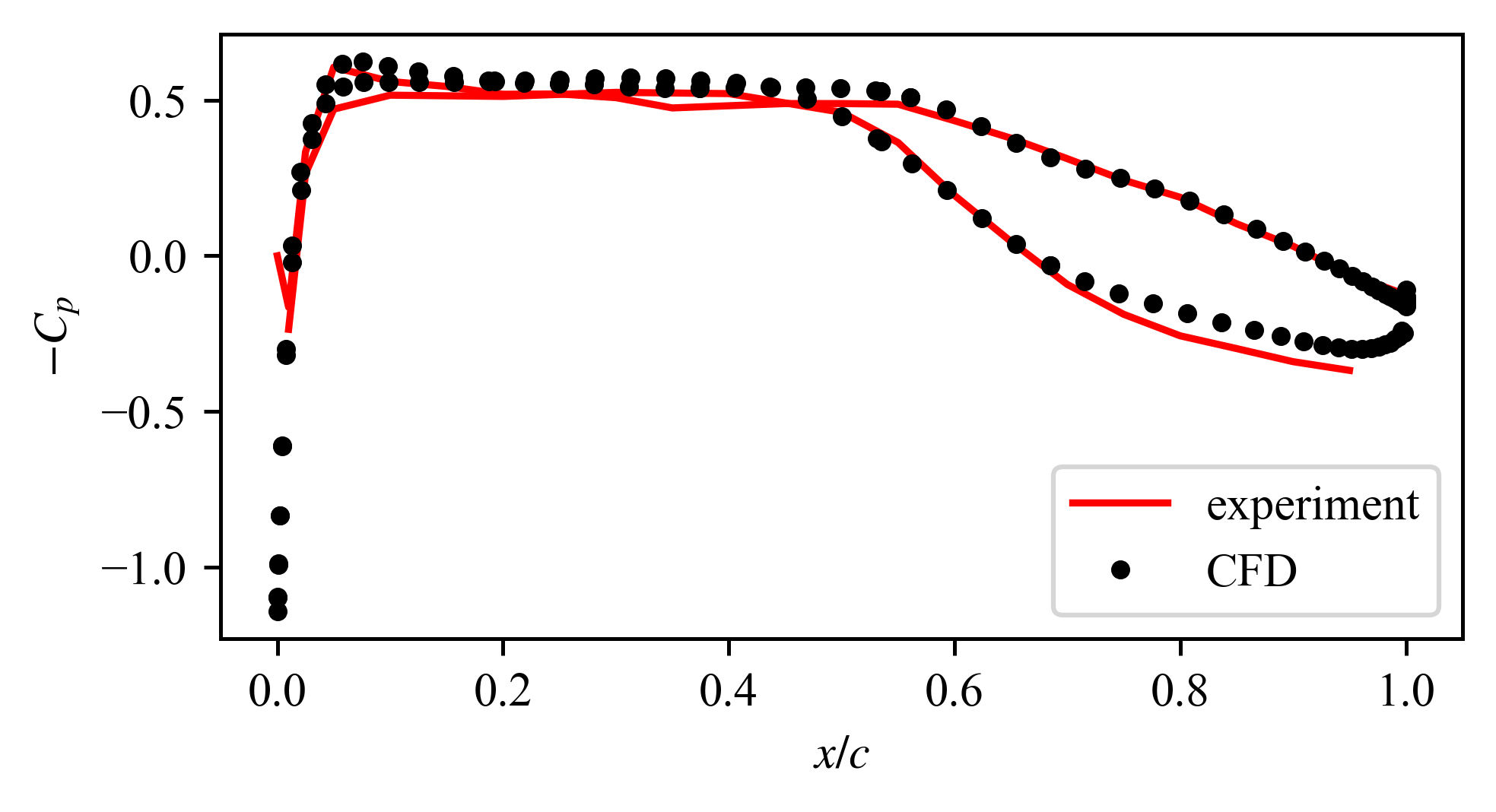}}
        \caption{Steady-state pressure coefficient at the 60\% span location for the static wing}
        \label{fig:case1val}
    \end{figure}

    Initial qualitative validation is conducted for the steady-state physics using the available experimental data~\cite{dansberry93,piatek03} for the rigid wing. Figure~\ref{valid1} presents the pressure coefficient at the 60\% span location for case 1 where it can be seen that there is reasonable agreement between the experiment and CFD, however, the shock is much stronger in the experiment. This could be resolved by mesh refinement, however, is outside of the scope of this work. Figure~\ref{valid2} compares the experimental and CFD pressure coefficient distributions at the 60\% span location for case 2a demonstrating good agreement. Table~\ref{tab:val} compares the experimental and full-order aeroelastic flutter conditions for $M_\infty = 0.74$, $\alpha_0 = 0^\circ$ and $M_\infty = 0.8$, $\alpha_0 = 0^\circ$ demonstrating that for both Mach numbers there is good agreement with experiment. 

    \begin{table}[h]
    \centering
    \begin{tabular}{ccccc}
        \hline
        &\multicolumn{2}{c}{$\bm{M_\infty = 0.74, \alpha_0 = 0^\circ}$} & \multicolumn{2}{c}{$\bm{M_\infty = 0.8, \alpha_0 = 0^\circ}$}\\
        \hline
        & $f_f$ [Hz] & $q_f$ [psf] &$f_f$ [Hz] & $q_f$ [psf]  \\
        \hline
        \hline
         Experiment~\cite{dansberry93} & 4.3 & 168.8  & 4.15 & 172  \\
         Present                        & 4.16 & 170 & 4.1 & 167 \\
         Error [\%] &  -3.25 & 0.71 & -1.21 & -2.91 \\
        \hline
    \end{tabular}
    \caption{Experimental validation of frequency and dynamic pressure}
    \label{tab:val}
    \end{table}

    \subsection{Forced Response}
    \label{sec:forced}
\begin{figure}[h]
        \centering
            \includegraphics[width=0.65\textwidth]{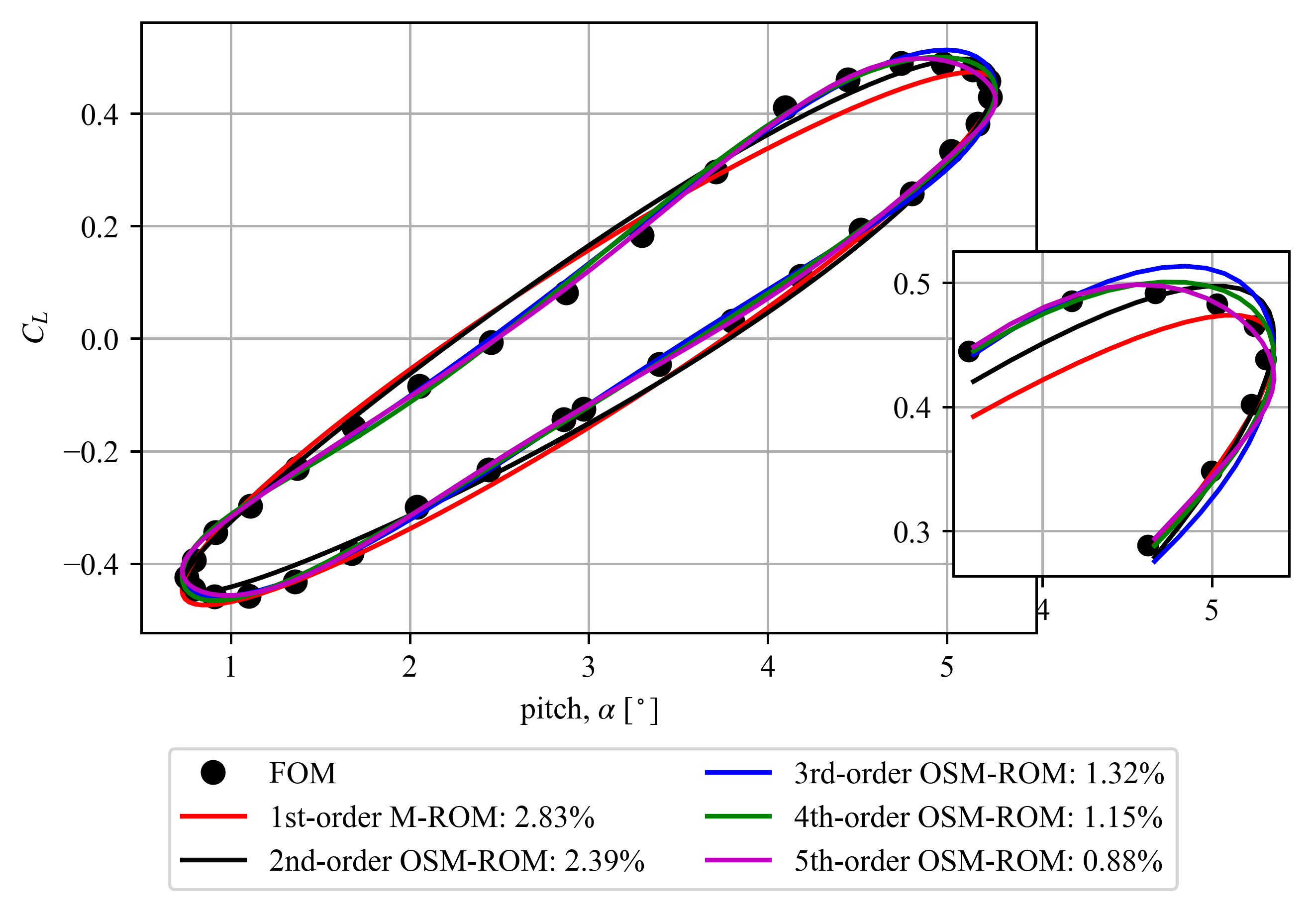}
        \caption{Lift coefficient as a function of rotation with forced sinusoidal motion for case 1}
        \label{fig:forced}
    \end{figure}

\begin{figure}[h]
        \centering
        
            \includegraphics[width=0.5\textwidth]{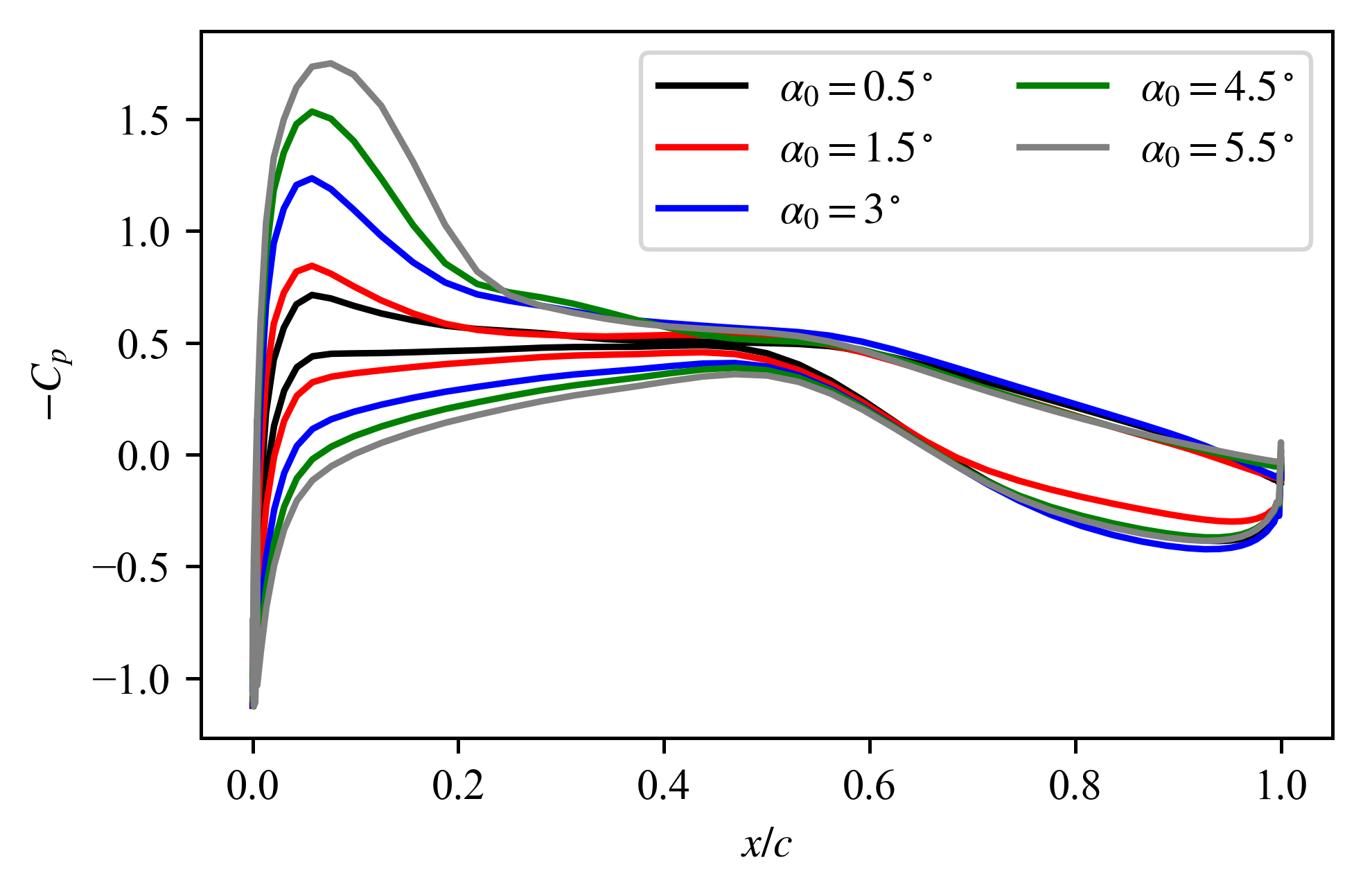}
        \caption{Pressure coefficient at various angles-of-attack spanning the range observed in the training data with $\bm{M_\infty = 0.7}$}
        \label{fig:forceCp}
    \end{figure}

    Figure~\ref{fig:forced} presents the unsteady lift force as a function of pitch angle for case 1 with the wing undergoing forced sinusoidal motion at $f_\alpha = 10$Hz, $|\alpha| = 2.25^\circ$. The inflections through the midpoints of the cycle, flattening/stretching at the extrema, and asymmetry are indicative of both odd- and even-order nonlinear shock wave behavior~\cite{candon19phd}. To probe further, the steady-state pressure coefficient at the 60\% span location is presented in Fig.~\ref{fig:forceCp}, confirming that the shock gets stronger and moves aft as the pitch angle increases, then disappears completely at the trough of the cycle. Returning To Fig.~\ref{fig:forced}, it is clearly important to include both the odd-order and even-order nonlinearities in the ROM. It can be seen that as the ROM order increases the accuracy improves. A minimum of third-order is required as to capture the general nonlinear form. Impressively, the fifth-order ROM is able to model the nonlinear unsteady aerodynamic forces with excellent precision, despite identifying only 48 of 80729 total terms (see Appendix~\ref{sec:append}) (which would require at least that many training samples and therefore weeks to train). 

    \clearpage
    
    \subsection{Steady-State Flowfields}

\begin{figure}[h!]
		\centering
        \subfigure[$-4^\circ$]{\label{}
			\includegraphics[width=0.32\textwidth]{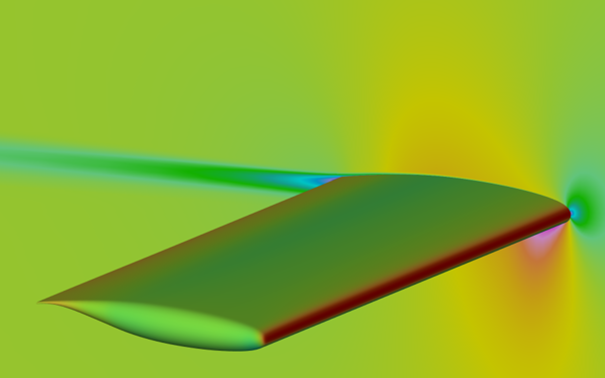}}
        \subfigure[$-2^\circ$]{\label{}
			\includegraphics[width=0.32\textwidth]{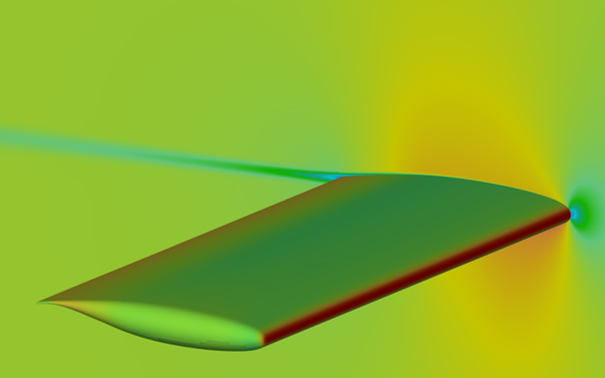}}
        \subfigure[$0^\circ$]{\label{}
			\includegraphics[width=0.32\textwidth]{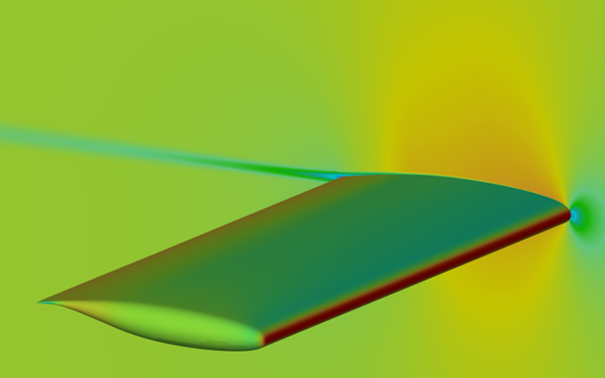}}
   
        \subfigure[$2^\circ$]{\label{}
			\includegraphics[width=0.32\textwidth]{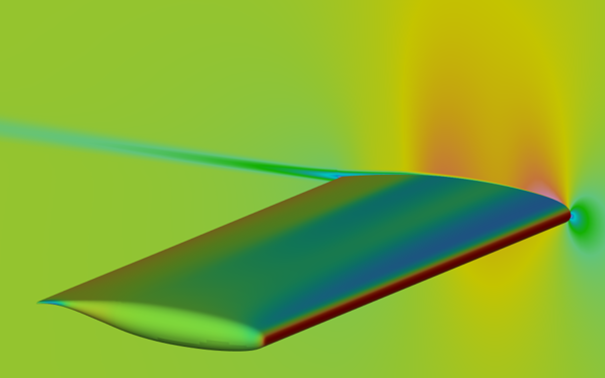}}
        \subfigure[$4^\circ$]{\label{}
			\includegraphics[width=0.32\textwidth]{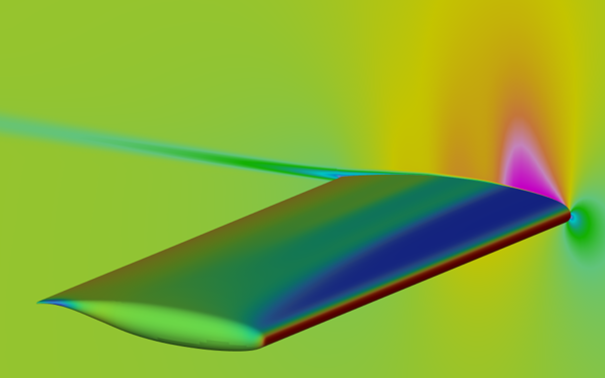}}
        \subfigure[$7^\circ$]{\label{}
			\includegraphics[width=0.32\textwidth]{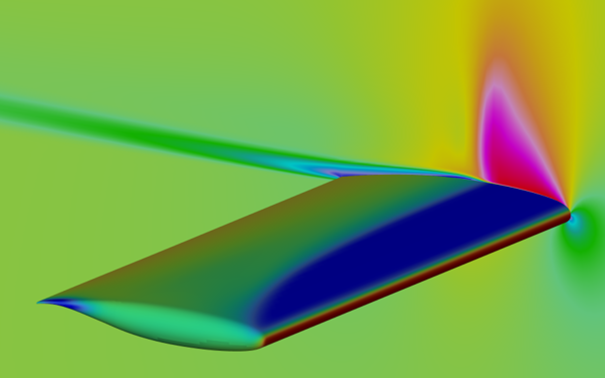}}
            \includegraphics[width=0.7\textwidth]{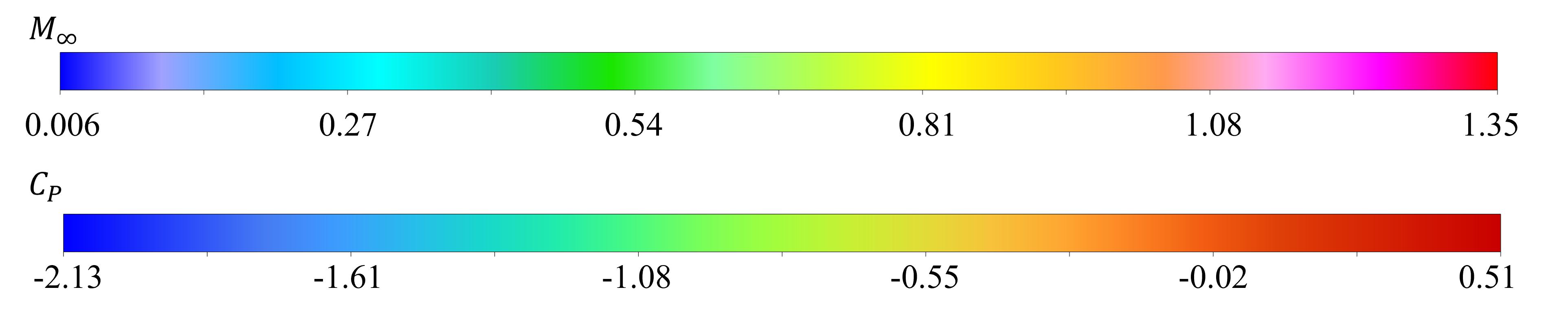}
		\caption{Steady-state flowfields depicting the pressure coefficient on the wing and Mach number on the symmetry plane for $M_\infty = 0.74$}
		\label{fig:flowfields_074}
	\end{figure}

    The steady-state flowfields spanning the range of maximum and minimum pitch amplitudes used to generate the training data for case 2a and 2b are presented in Fig.~\ref{fig:flowfields_074}. It can be seen that as the pitch amplitude increases to $\alpha_0 = 2^\circ$ a weak double shock appears on the upper surface, increasing in strength and moving aft as $\alpha_0$ increases. At $\alpha_0 = 7^\circ$ a strong shock is present which is characterized by some boundary layer separation downstream. Furthermore, 3D effects can be clearly observed. Although not shown here, on the lower surface a weak shock appears at $\alpha_0 = -2^\circ$, increasing in strength and moving aft. At $\alpha = -4^\circ$ the shock is stronger and a moderate separation region can be observed downstream. Noting that the training data for case 2a spans $\alpha_0 = -4^\circ \ \mathrm{to} \ 2^\circ$, it is characterized by mildly nonlinear unsteady transonic behavior, including the appearance and disappearance of shock waves on both surfaces and some separation. For case 2b for which the training data spans $\alpha_0 = 1^\circ \ \mathrm{to} \ 7^\circ$, the problem is characterized by variable strength and disappearance of the upper surface shock and strong 3D effects, while no shock appears on the lower surface. 

\begin{figure}[h!]
		\centering
        \subfigure[$0^\circ$]{\label{}
			\includegraphics[width=0.32\textwidth]{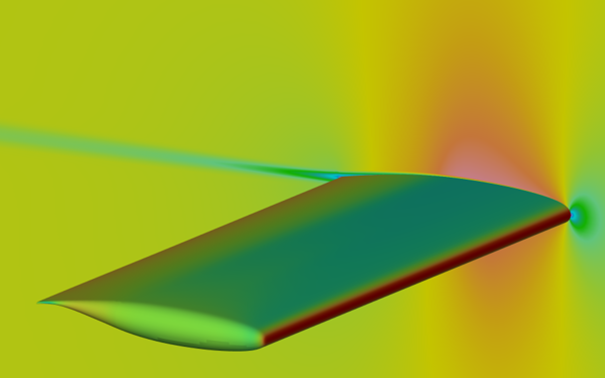}}
        \subfigure[$2^\circ$]{\label{}
			\includegraphics[width=0.32\textwidth]{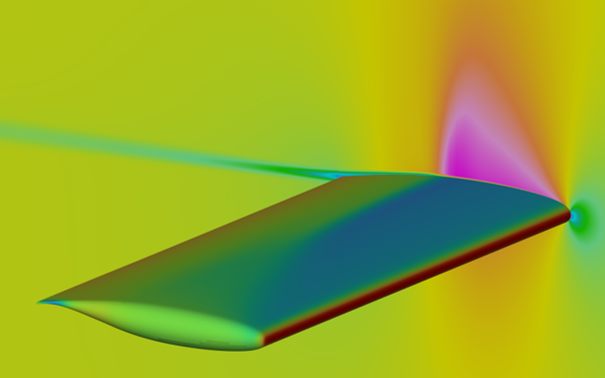}}
        \subfigure[$3^\circ$]{\label{}
			\includegraphics[width=0.32\textwidth]{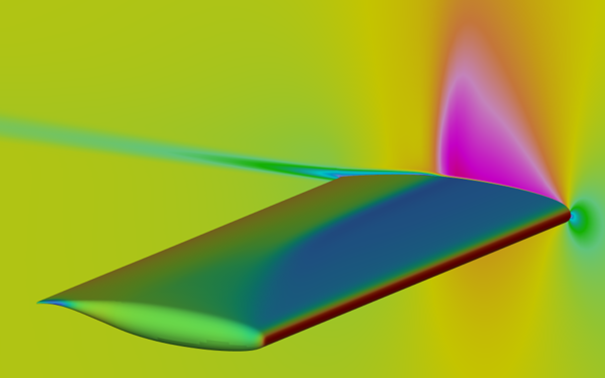}}
   
        \subfigure[$4^\circ$]{\label{}
			\includegraphics[width=0.32\textwidth]{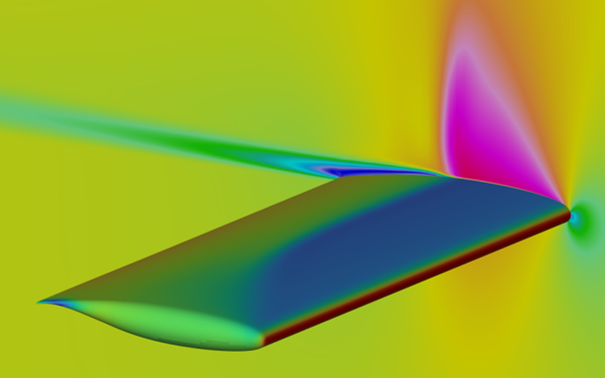}}
        \subfigure[$5^\circ$]{\label{}
			\includegraphics[width=0.32\textwidth]{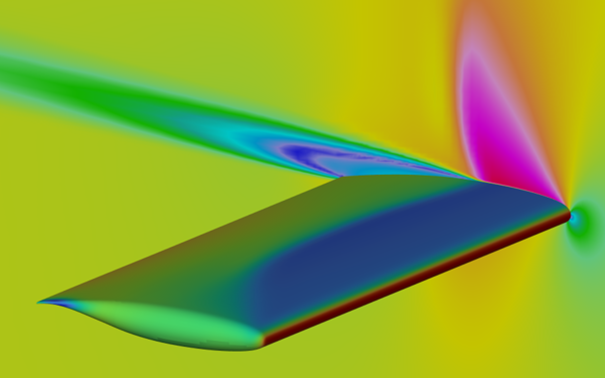}}
        \subfigure[$6^\circ$]{\label{}
			\includegraphics[width=0.32\textwidth]{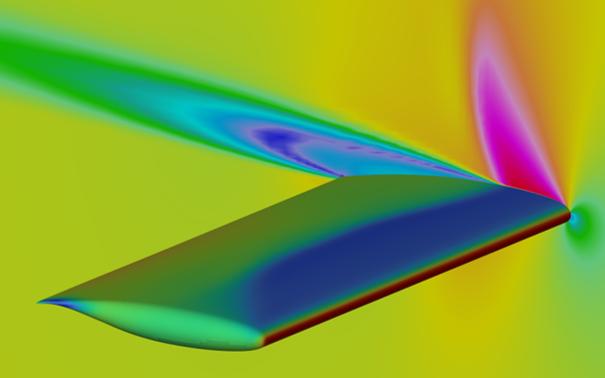}}
            \includegraphics[width=0.7\textwidth]{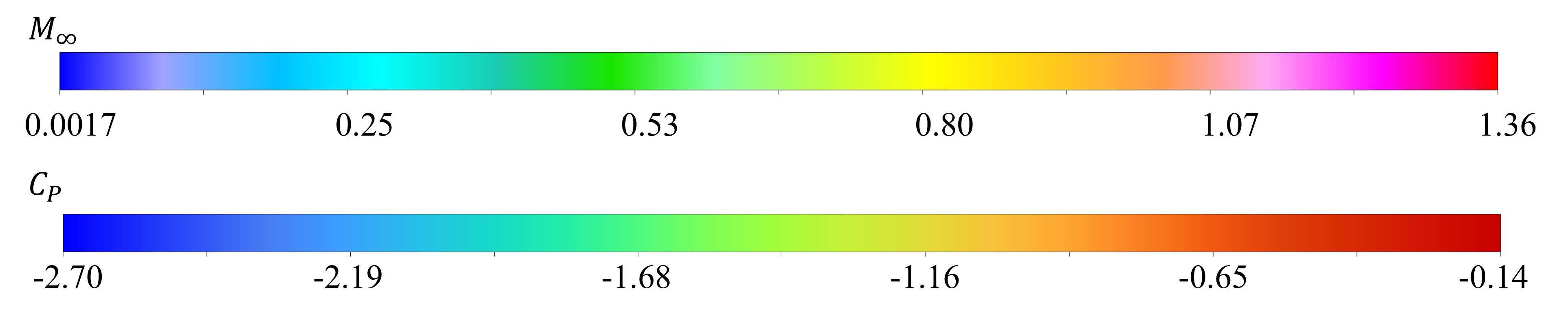}
		\caption{Steady-state flowfields depicting the pressure coefficient on the wing and Mach number on the symmetry plane for $M_\infty = 0.8$}
		\label{fig:flowfields_08}
	\end{figure}

    The steady-state flowfields spanning the range of maximum and minimum pitch amplitudes used to generate the training data for case 3a and 3b are presented in Fig.~\ref{fig:flowfields_08}. Here the nonlinear transonic aerodynamic effects are more pronounced. As the AoA increases, the upper surface shock increases in strength, beginning aft of the elastic axis and moving across it towards the leading edge. Above $\alpha_0 = 4^\circ$ significant boundary layer separation and a large recirculation region can be observed. With a portion of the training data/aeroelastic response characterized by such unsteady flow, it may mean that the fundamental assumptions associated with nonlinear time-invariant model reduction are violated. Therefore, this Mach number provides a robust assessment of the proposed method. It should be noted that global flow instabilities such as shock buffet are unlikely given coarse spatial and temporal resolution considered. 
    
   


    \clearpage
    \subsection{Time-Accurate Flutter}
    \label{sec:flutter}

    In this section the time-accurate flutter results are presented for ROMs of increasing order. An important distinction to make is that the first-order ROM presented here is not equivalent to the linearized impulse response function of the system. Rather, it is a first-order approximation to the nonlinear training data. This is not necessarily the \enquote{best} linear model that can be derived in terms of linear flutter predictions. Accordingly, even for cases where it would be expected that a linearized model should match the flutter speed of the FOM very closely, such performance of the first-order ROM here is not guaranteed. As such, the approach proposed in this paper is recommended as an efficient means of generating nonlinear models that can accurately model both linear and nonlinear aeroelastic behavior. Recommendations to improve the base linear model are provided in the conclusion.

\begin{figure}[h]
        \centering
        \subfigure[Case 2a: $M_\infty = 0.74, \ \alpha_0 = 0^\circ$]{\label{case2a}
            \includegraphics[width=0.47\textwidth]{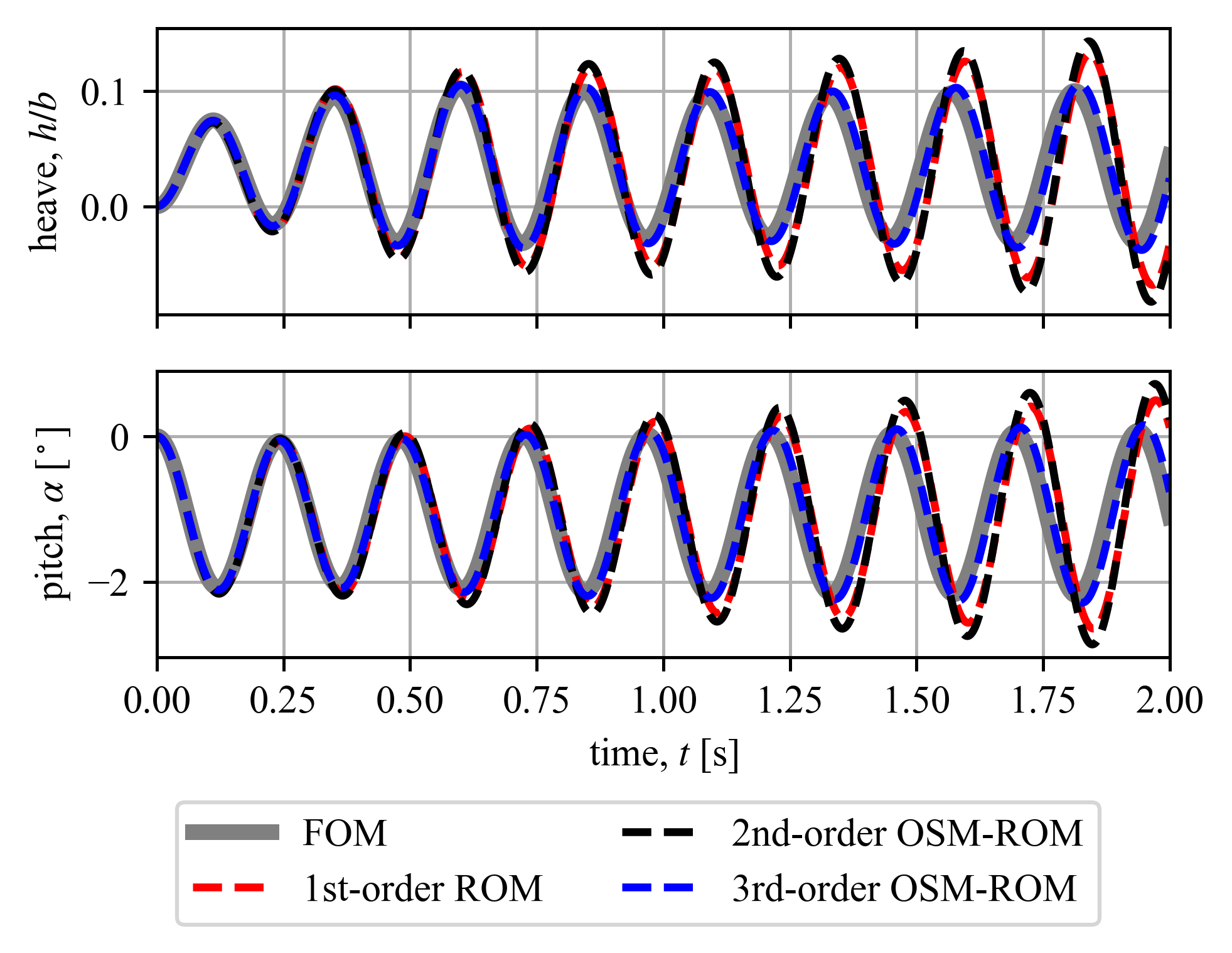}}
        \subfigure[Case 2b: $M_\infty = 0.74, \alpha_0 = 3^\circ$]{\label{case2b}
            \includegraphics[width=0.47\textwidth]{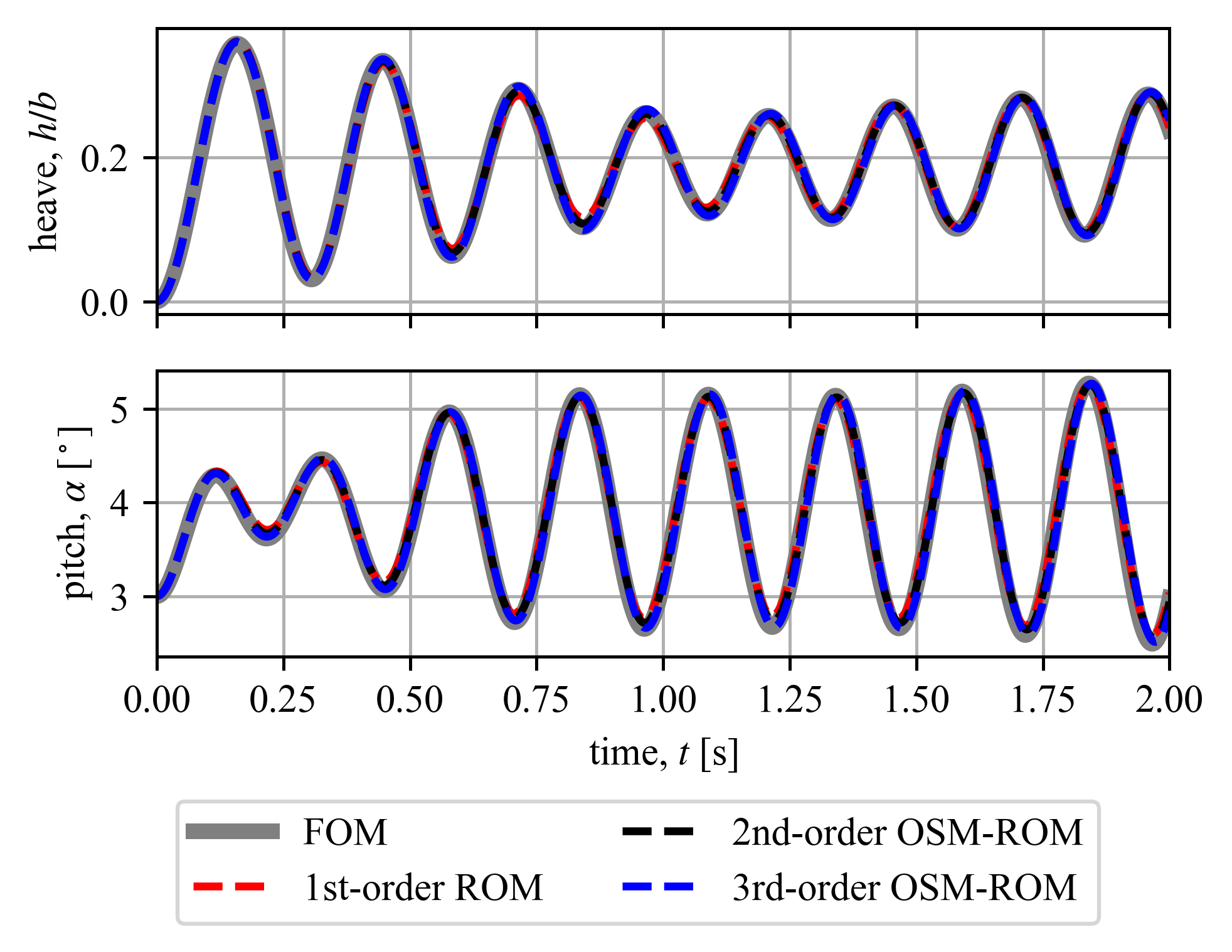}} 
        \caption{Time responses comparing the FOM and various ROM solutions for case 2 at $q_\infty=174.4$}
        \label{fig:case2a_flutter_time}
    \end{figure}

    Table~\ref{tab:flutter} presents the flutter results for all cases comparing the FOM and various ROM solutions. For case 2a the first-order and second-order approximations are conservative, under predicting the flutter dynamic pressure by more than 4\%, while the third-order OSM-ROM under predicts by 0.4\%, an order of magnitude improvement in accuracy. The flutter frequency is captured with good accuracy by all ROMs relative to the FOM. The time-domain solution at a dynamic pressure of $q_\infty = 174.4$ psf (slightly above the flutter speed) is presented in Fig.~\ref{case2a}. It can be seen that the third-order OSM-ROM captures the response with good accuracy, while the first-order ROMs and second-order OSM-ROM both display rapidly diverging responses. For case 2b all ROMs can be seen to perform very well, also observable in Fig.~\ref{case2b} which presents the time responses at a dynamic pressure of $q_\infty = 174.4$ psf for case 2b.

\begin{figure}[h]
        \centering
            \includegraphics[width=1\textwidth]{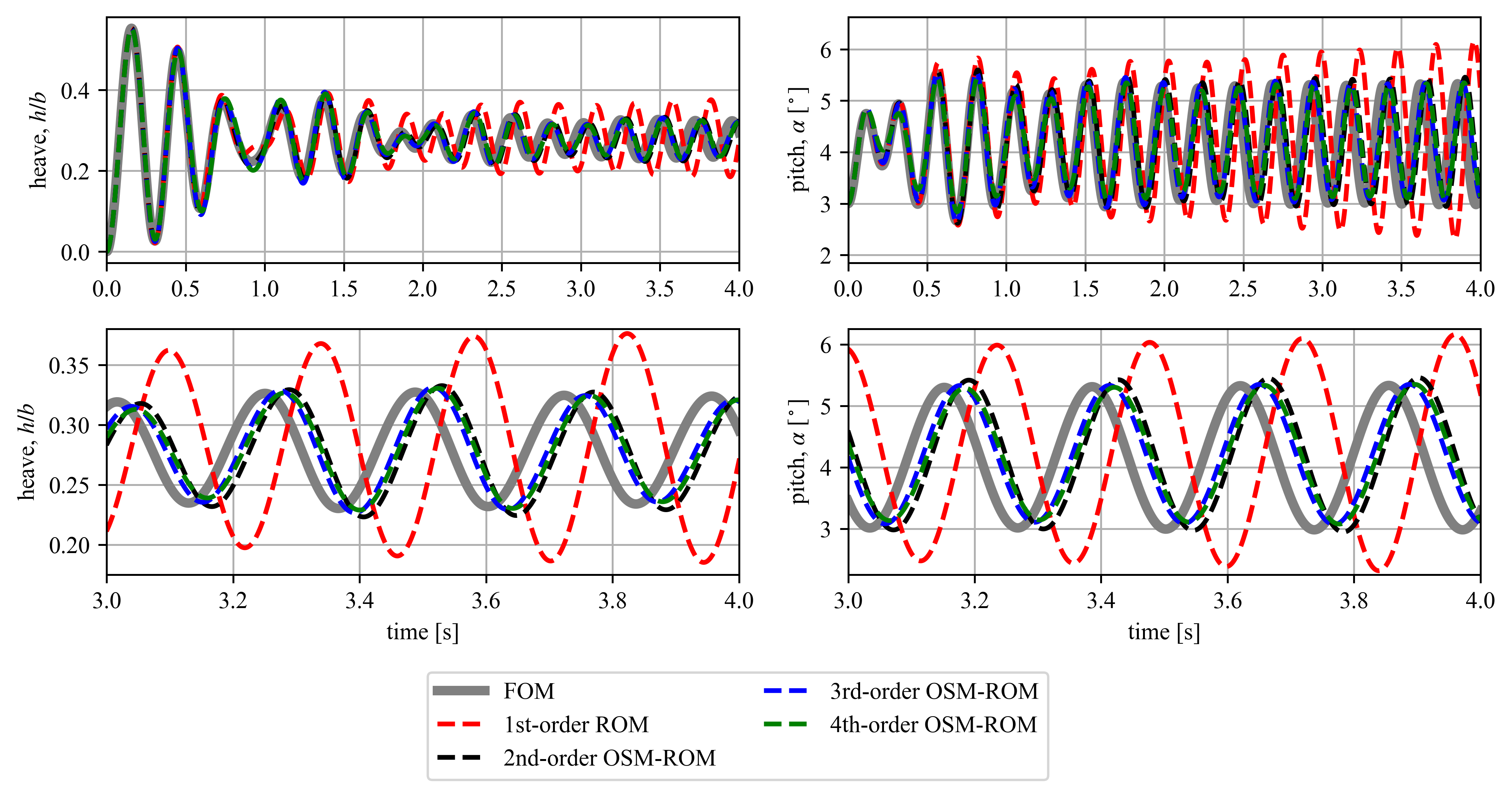}
        \caption{Time response comparing the FOM and various ROM solutions at $q_\infty = 251$ psf for case 3b: $M_\infty = 0.8, \ \alpha_0 = 3^\circ$}
        \label{fig:case3b_flutter_time}
    \end{figure}

    For case 3a the first-order and second-order ROMs over predict the flutter speed by 10\% and 8\% respectively. The third-order ROM improves the prediction significantly, over predicting by 1.6\%. The fourth-order ROM provides an under prediction of 0.17\%.  Surprisingly, for case 3b the first-order and second-order ROMs are able to predict the flutter speed with greater accuracy than for case 3a with an under-prediction of 7.3\% and 1.8\% respectively. The third-order and fourth-order ROMs both provide an excellent prediction of the flutter speed  relative to the FOM, although the identification of the fourth-order ROM does not offer any notable improvement for this case. The time responses at a dynamic pressure of $q_\infty = 251$ psf (slightly above the FOM flutter speed) are presented in Fig.~\ref{fig:case3b_flutter_time}. The divergence of the first-order ROM solution becomes noticeable after approximately 0.5 s and is pronounced after approximately 2 s, clearly highlighting the need for nonlinear terms. 
    
\begin{figure}[h]
        \centering
            \includegraphics[width=1\textwidth]{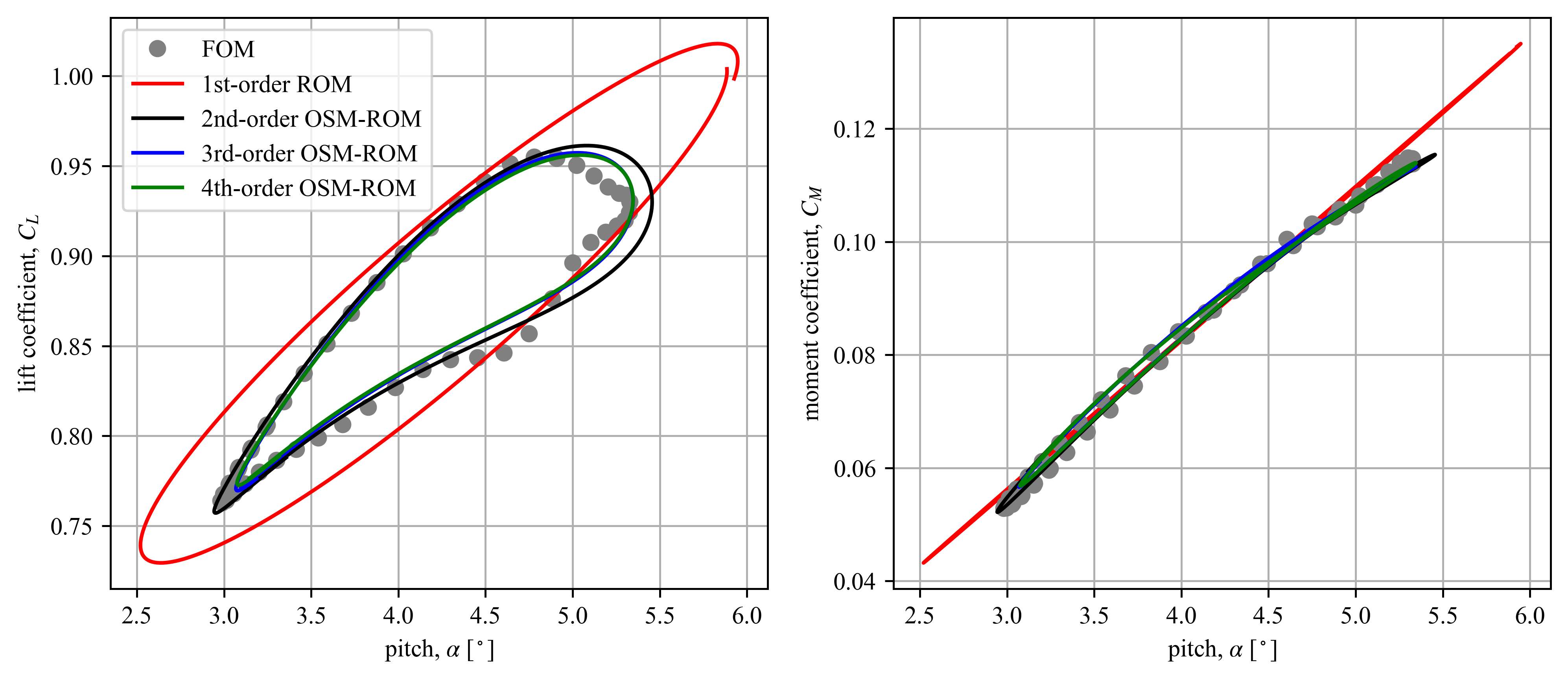}
        \caption{Force versus pitching motion for the FOM and various ROM solutions at $q_\infty = 251$ psf for case 3b: $M_\infty = 0.8, \ \alpha_0 = 3^\circ$}
        \label{fig:case3b_flutter_force}
    \end{figure}

    Although the nonlinear models provide an excellent prediction of the flutter speed, phase errors are present. This is unsurprising given the complexity of the nonlinear aerodynamic forces at this condition which can be observed in Fig.~\ref{fig:case3b_flutter_force}. At the peak of the cycle, the effects of stall and boundary layer separation can be observed, through a phase lead in lift coefficient which can not be predicted by the first-order ROM. The nonlinear ROMs do a reasonable job at approximating the nonlinear aerodynamic forces, capturing the general form well, although detailed features of the nonlinear forces related to stall, separation and re-attachment cannot be modeled by this class of ROM. The phase lead is captured by the nonlinear approximation although it is less so than the FOM, hence, explaining phase error in the response.

    A summary of the linear and nonlinear terms identified for each of these cases is provided in the Appendix~\ref{sec:append}. Firstly the total number of terms identified is noteworthy, where less than fifty of the tens-of-thousands of possible terms are all that is required - a promising finding for the future of nonlinear aeroelastic modeling. Secondly, the distribution of linear and nonlinear terms is informative. It can be seen that for case 2a a significant portion of the nonlinear terms are cross-terms, while for case 2b no nonlinear cross-terms are identified. For case 3a, nearly half of the nonlinear terms for the heave mode are cross-terms, while for the pitch mode the portion of cross-terms is much smaller. An opportunity exists to study the sparsity patterns further and how they relate to the unsteady aerodynamic / aeroelastic physics being observed.



    \begin{table}[h]
    \centering
    \begin{tabular}{cccccccc}
        \hline
         && FOM & $1^{st}$-order & $2^{nd}$-order& $3^{rd}$-order& $4^{th}$-order \\
        \hline
        \hline
         \multirow{ 2}{*}{case 2a} &$f_f$ [Hz]  & 4.16  & 4.1 & 4.1 & 4.15 & - \\
         &$q_f$ [psf]  & 169.4 & 162.5 & 162.6 & 168.5 & - \\
         \hline
         \multirow{ 2}{*}{case 2b} & $f_f$ [Hz] & 4.0 &  4.0 & 4.0 & 4.0 & -\\
         & $q_f$ [psf]  & 151.6 & 150.8 & 150.9 & 151.2 & -  \\
        \hline
        \multirow{ 2}{*}{case 3a} &$f_f$ [Hz]  & 4.10 & 4.10 & 4.10 & 4.10 & 4.10 \\
         &$q_f$ [psf]  & 168.1 & 183.9 &    181.6 & 170.8 & 167.8 \\
         \hline
         \multirow{ 2}{*}{case 3b} & $f_f$ [Hz] & 4.21 &  4.23 & 4.21 & 4.21 & 4.21\\
         & $q_f$ [psf]  & 249.6 & 231.4 & 245.2 & 250.6 & 250.8  \\
        \hline
    \end{tabular}
    \caption{Flutter frequency and dynamic pressure for all cases}
    \label{tab:flutter}
    \end{table}

    \begin{figure}[h]
        \centering
            \includegraphics[width=0.5\textwidth]{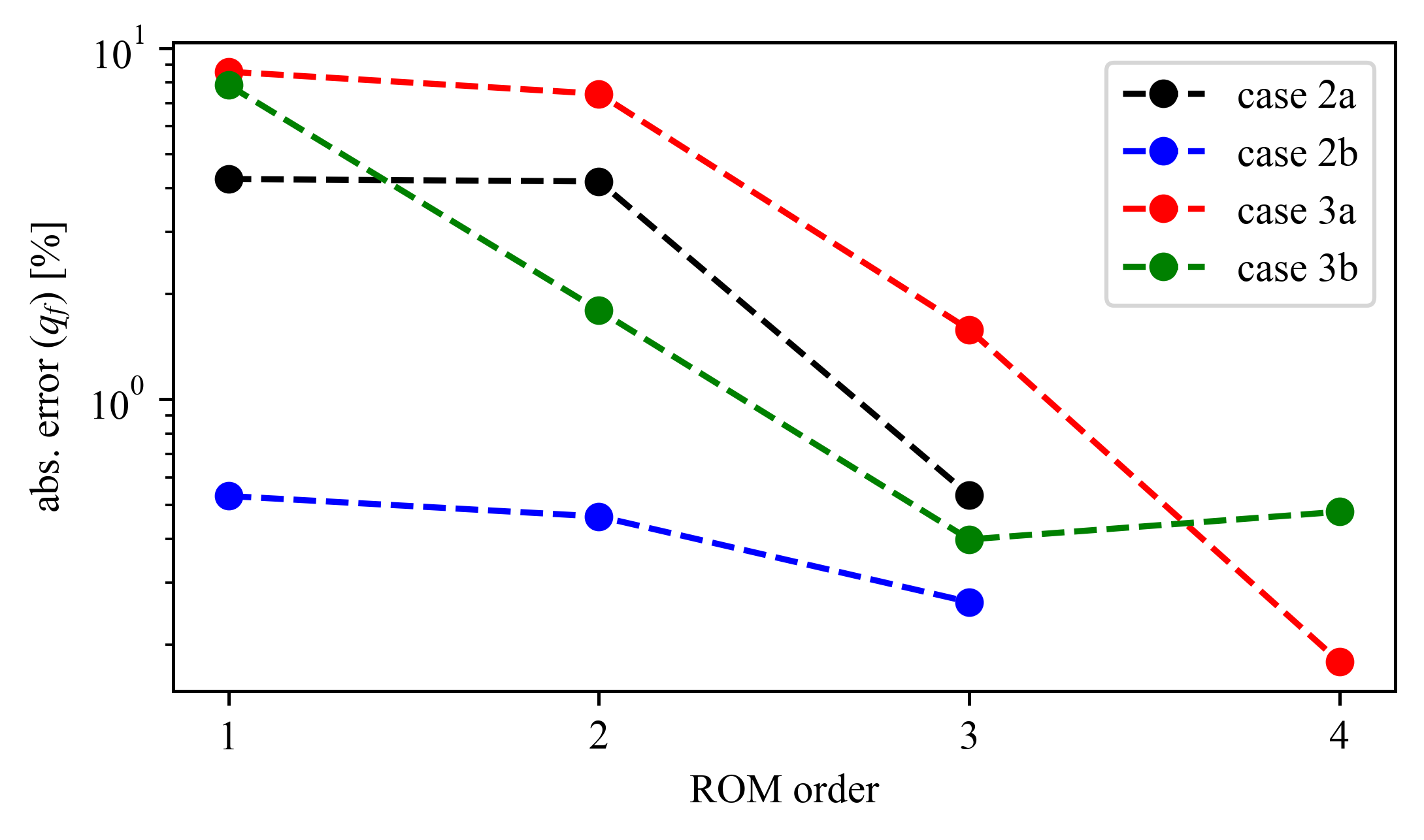}
        \caption{Absolute error in ROM flutter dynamic pressure predictions}
        \label{fig:flutter_error}
    \end{figure}

    \clearpage

    \subsection{Limit Cycle Oscillation}
\begin{figure}[h]
        \centering
        \subfigure[Case 2b: $M_\infty = 0.74, \ \alpha_0 = 3^\circ$]{\label{2blco}
            \includegraphics[width=1\textwidth]{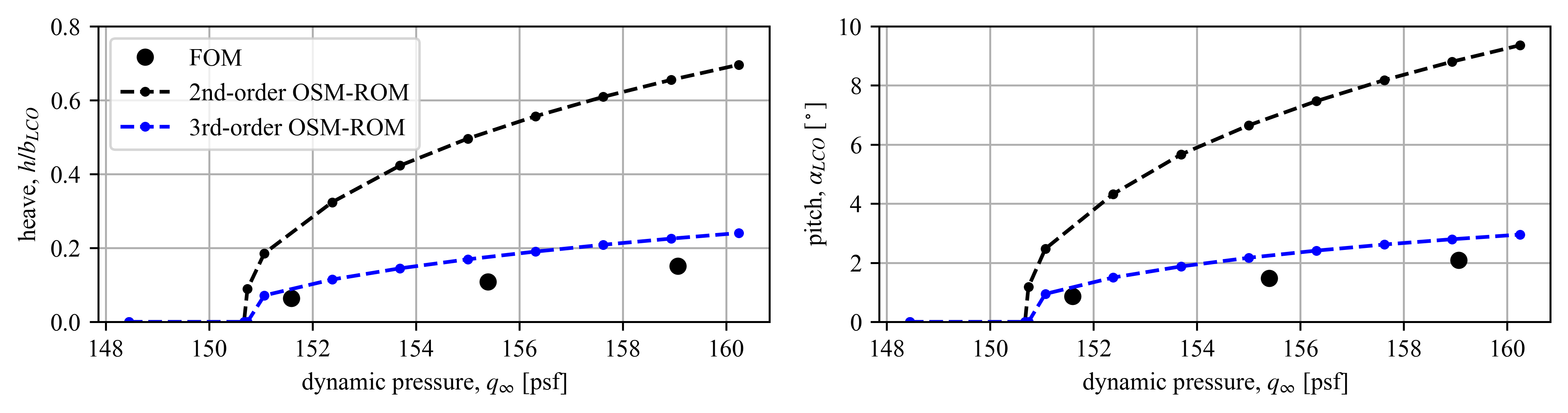}}
            
            \subfigure[\textbf{Case 3a: $M_\infty = 0.8, \ \alpha_0 = 2^\circ$}]{\label{3alco}
            \includegraphics[width=1\textwidth]{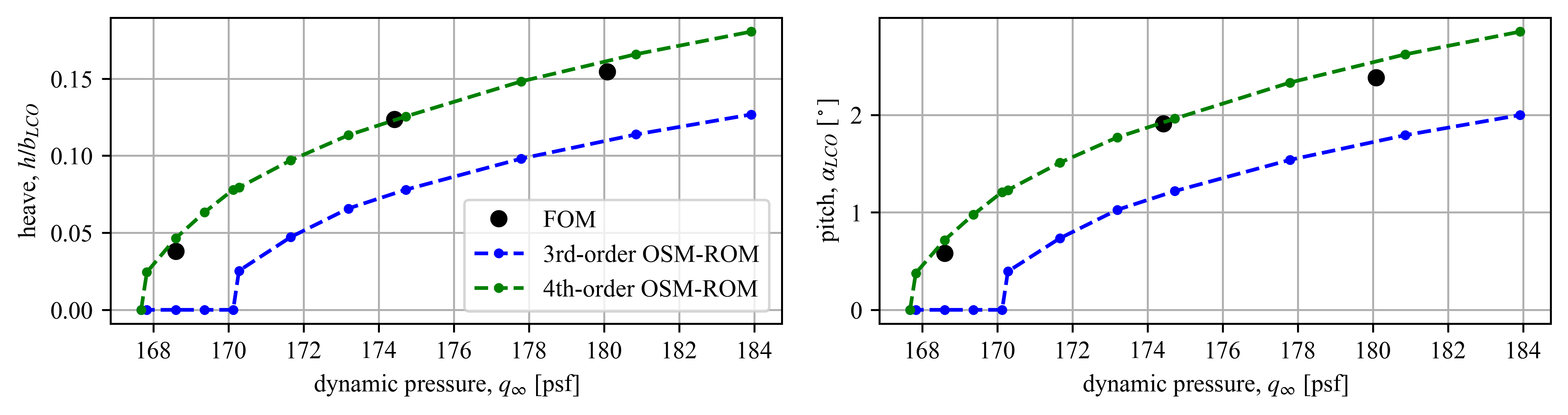}}
        \caption{Post flutter LCO amplitude for case 2b and case 3a}
        \label{fig:case2b_LCO}
    \end{figure}

    Limit cycles are found for case 2b and case 3a conditions only. Figure~\ref{2blco} presents the LCO amplitude for post flutter dynamic pressures for case 2b. Both the second-order and third-order OSM-ROMs predict LCO, however, the error in the second-order approximation is very large. Higher-order approximations (above order 3) offer no improved accuracy for this case. The sparse third-order OSM-ROM performance is generally good - better in the pitch mode than heave. The performance is excellent close to the flutter point and degrades deeper into the post-flutter region. This supports the findings of Section~\ref{sec:forced}, suggesting that in order to accurately capture the nonlinear transonic flow physics associated with LCO, both even- and odd-ordered terms are needed. 
    
    The LCO for case 3a is characterized by more pronounced nonlinear flow features, such as, large scale shock wave motion and boundary layer separation - as was presented in Fig.~\ref{fig:flowfields_08}. For this case, only the third-order and fourth-order approximations predict LCO, presented in Fig.~\ref{3alco}. The third-order model under predicts the LCO amplitude, while the inclusion of 4th-order terms allows the LCO to be modeled with high accuracy. It appears that the reduced accuracy of the third-order model is predominantly related to the over prediction of the flutter speed, rather than an inability to model the nonlinear physics.

\begin{figure}[h]
        \centering
        \subfigure[heave]{\label{aa}
            \includegraphics[width=1\textwidth]{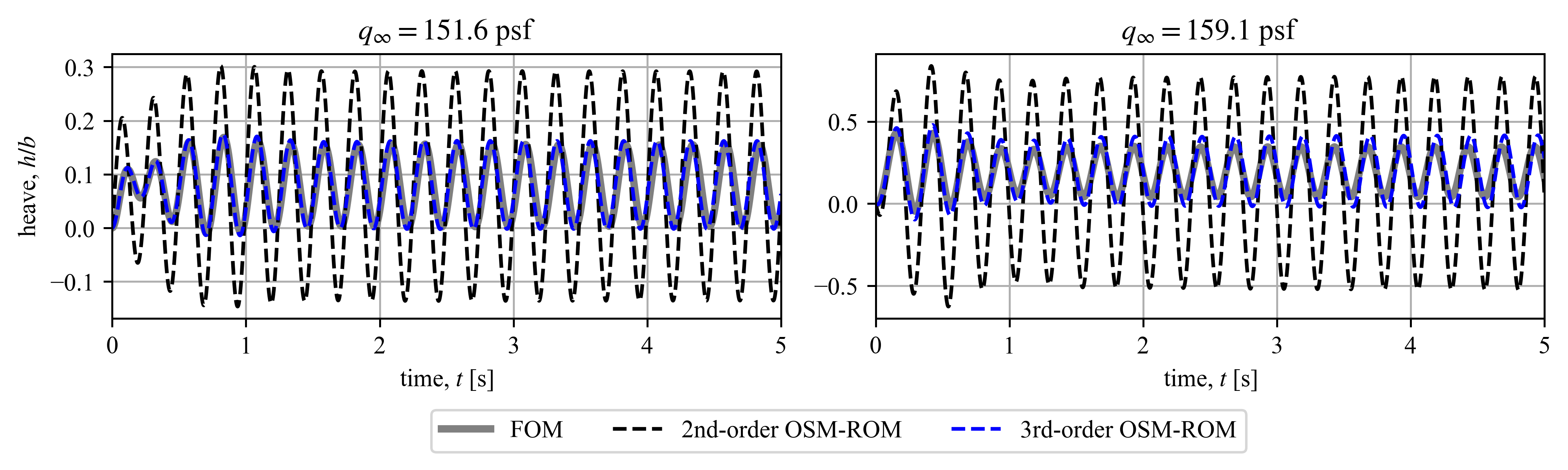}}

        \subfigure[pitch]{\label{aa}
            \includegraphics[width=1\textwidth]{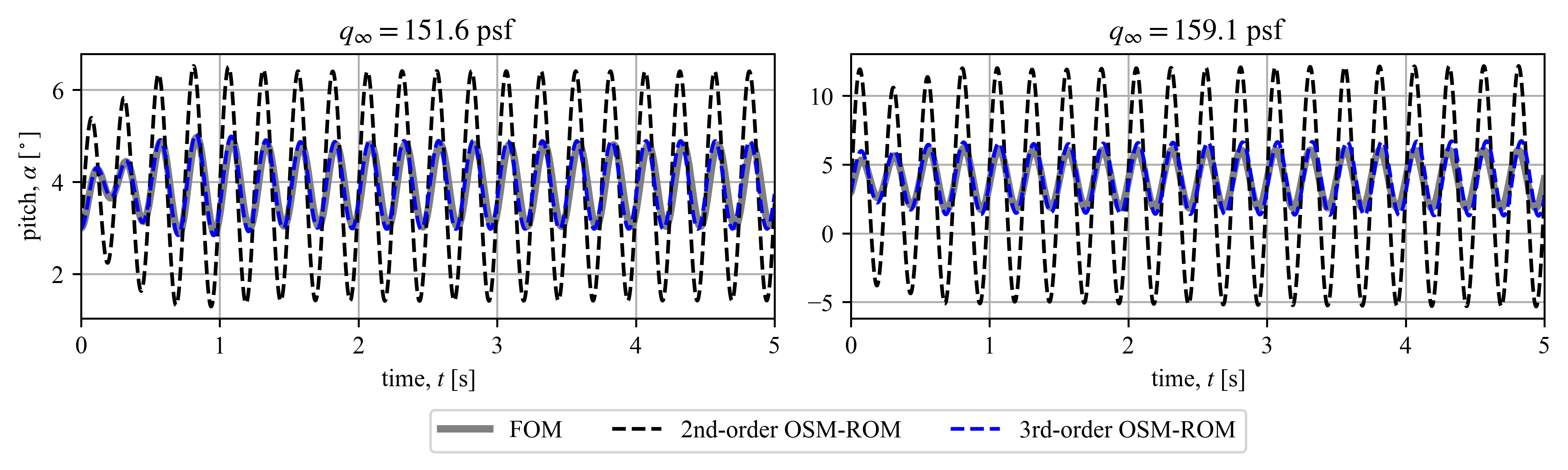}}
        \caption{Time response of the LCO for case 2b: $M_\infty = 0.74, \ \alpha_0 = 3^\circ$}
        \label{fig:case2b_time}
    \end{figure}

    Figure~\ref{fig:case2b_time} presents the time histories at the lowest and highest FOM dynamic pressures for case 2b. It can be seen that the third-order OSM-ROM captures the LCO with reasonable precision, including in phase, for all cases, while there is an over prediction of the amplitude for the highest dynamic pressure. 

\begin{figure}[h]
        \centering
            \includegraphics[width=1\textwidth]{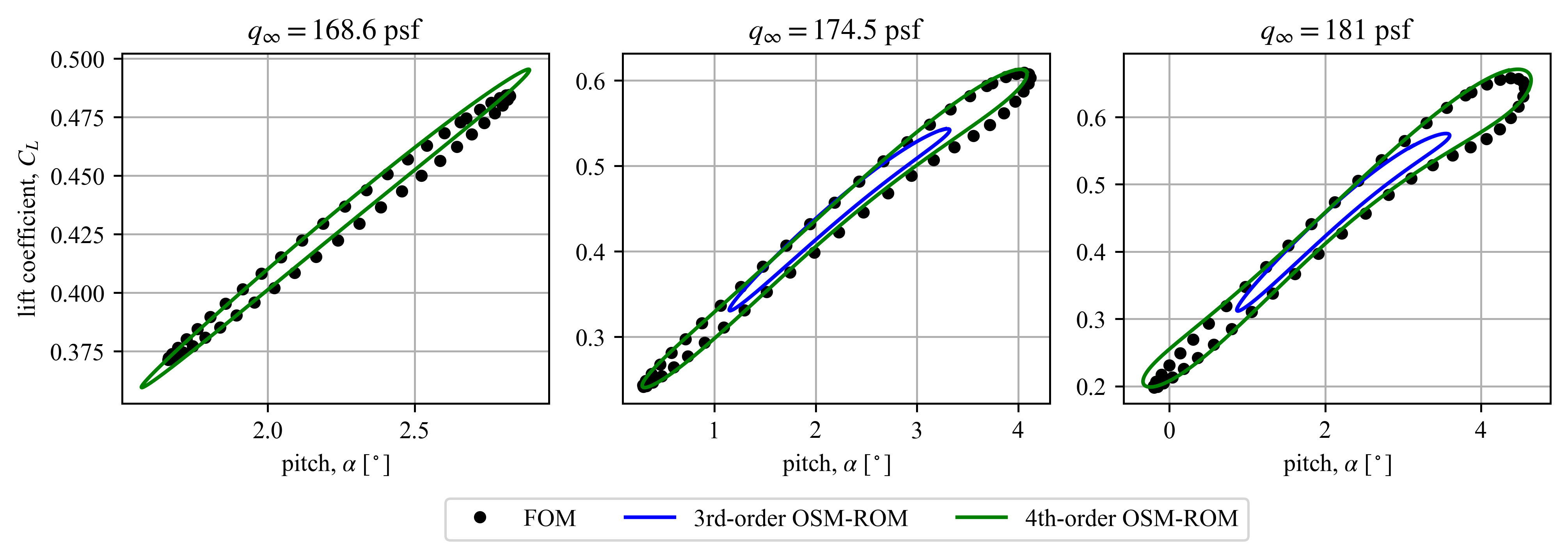}
        \caption{Lift force versus pitching motion for various limit cycles for case 3a: $M_\infty = 0.8, \ \alpha_0 = 2^\circ$}
        \label{fig:case3b_clalpha}
    \end{figure}

    In Fig.~\ref{fig:case3b_clalpha} the relationship between pitching motion and the unsteady lift coefficient (generalized force in the heave mode) is presented for different LCO amplitudes for case 3a. For the lowest amplitude ($q_\infty = 168.6$ psf) a seemingly linear distribution can be observed where the forces and displacements are in phase. For this case, only the FOM and fourth-order OSM-ROM predict LCO. The general form is predicted well, however, there is an over prediction of force and displacement magnitude. This can be attributed to the fact that close to the flutter speed, small errors in the predicted flutter speed can lead to significant error in the prediction of the LCO amplitude.  As the LCO amplitude increases ($q_\infty = 174.5$ psf) a phase lead in force can be observed as the curve broadens which is captured well by the fourth-order OSM-ROM. For the largest amplitude ($q_\infty = 181$ psf) the force phase lead is increased and the qualitative nonlinear features of the curve become more pronounced. On a whole the LCO is predicted well by the fourth-order OSM-ROM, however, a discrepancy can be observed at the trough of the cycle. The fourth-order OSM-ROM predicts a force phase lead (similar to at the cycle peak), which does not exist for the FOM. This can be explained by incomplete training data, $i.e.$, the trough of the cycle is at a negative pitch angle which is outside the range of the training data for this case which spans 0$^\circ$ to 5$^\circ$. 

\begin{figure}[h]
        \centering
            \includegraphics[width=1\textwidth]{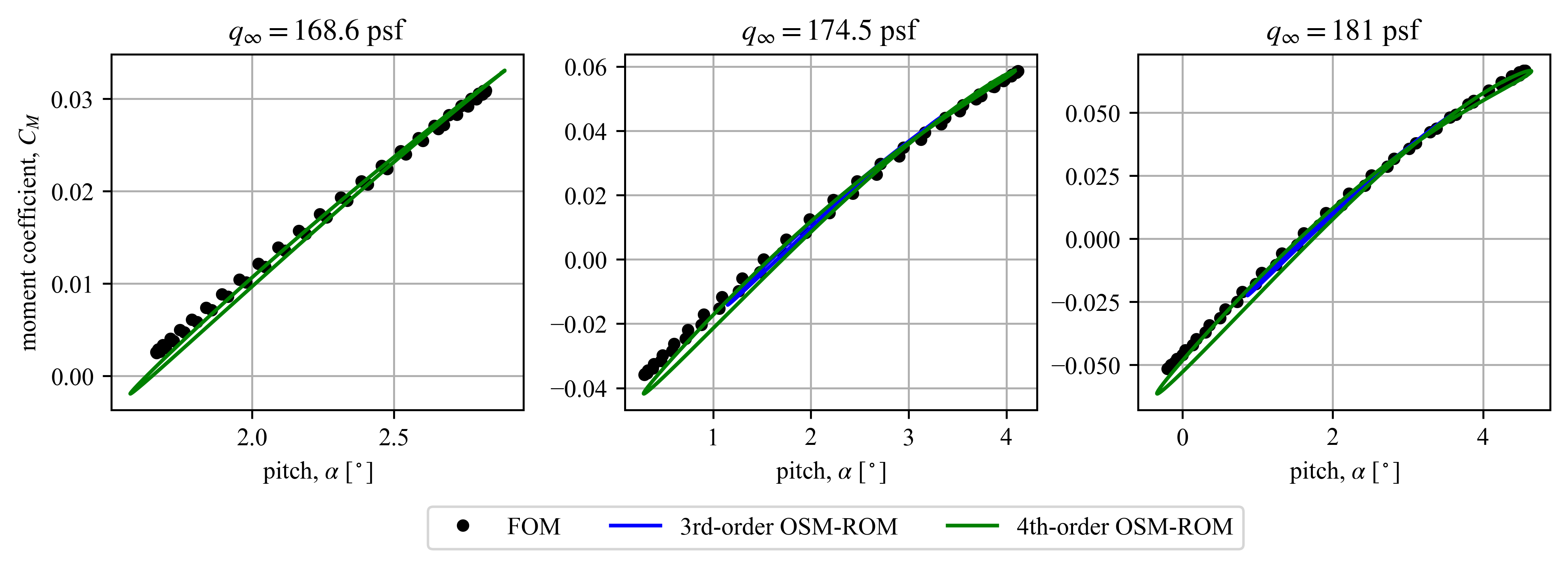}
        \caption{Pitching moment versus pitching motion for various limit cycles for case 3a: $M_\infty = 0.8, \ \alpha_0 = 2^\circ$}
        \label{fig:case3b_cmalpha}
    \end{figure}

    Similarly, in Fig.~\ref{fig:case3b_cmalpha} the relationship between pitching motion and the unsteady moment coefficient (generalized force in the pitch mode) is presented for different LCO amplitudes. Here the even-ordered nonlinearity in the system is clearly evident as the LCO amplitude grows, $i.e.$, directly related to the large-scale rocking motion of the upper-surface shock, which also traverses the elastic axis somewhere between $\alpha_0 = 4^\circ$ and $5^\circ$. This may provide some rationale as to the significant improvement in performance achieved using a fourth-order model. Again it can be seen that the peak of the cycle is very well predicted by the fourth-order OSM-ROMs, while there is error at the trough which is outside the range of the training data.  
    
    It is quite a remarkable finding that these ROMs are able to provide such an accurate prediction of the limit cycle behavior despite only a small fraction of nonlinear terms being identified (summarized in Section~\ref{sec:append}). For example, the third-order OSM-ROM for case 2b has two and four third-order direct terms only in the pitch and heave modes respectively, and no cross terms. This suggests that large complex nonlinear models are not necessary to model complex nonlinear transonic flow phenomena in an aeroelastic setting. This is of major significance for nonlinear aeroelastic model reduction and the future development of aeroelastic flight simulator technology. 

    \clearpage

    \subsection{Computational Savings}
    To ensure a stable limit cycle is achieved, at least 20,000 numerical time-steps are required without an initial perturbation. In this paper, an initial perturbation was used to reduce the number of time-steps to 5000. Given that the size of the perturbation was determined using the ROM, the online savings are computed based on 20,000 time-steps. For flutter the number of time-steps computed is 2000-4000.  Any full-order solutions are run on an Intel Xeon Gold 6152 CPU using 70 cores, while any ROM solutions on a single core ($i.e.$, insignificant resources). One FOM simulation to find LCO takes approximately 5 days, while ROM simulation takes seconds. 

    Of greater significance is the offline computational savings that are achieved by employing sparsity promotion. The maximum number of terms for a single ROM in this paper was 80,729 which would have required at least as many CFD-based training samples and weeks to generate. Rather, given OMP-based optimal coefficient selection, a maximum of 3000 training samples are used (half for training and half for cross-validation), requiring approximately 6-18 hours to compute and without sacrifice to fidelity. Theoretically the number of training samples could be reduced further given that less than 50 terms need to be identified. The difficulty is ensuring that all frequencies, amplitudes, and nonlinear modal interactions are adequately represented by the training set. There is a strong recommendation to explore this further.

    \section{Summary and Conclusion}
    \label{sec:conc}

    This paper proposes a novel nonlinear aeroelastic ROM framework based on sparse multi-input polynomial functionals that is robust, accurate, simple, and has an offline computational cost that is larger but comparable to that associated with training linearized ROMs. 

    The ROM framework utilizes orthogonal matching pursuit (OMP) to automatically select a sparse set of nonlinear multi-input aerodynamic coefficients from input (generalized displacements) and output (generalized aerodynamic force) relations. Through OMP the number of coefficients identified for all ROMs presented in this paper is less than 50 of tens-of-thousands of possible coefficients, reducing the offline computational costs by at least two orders of magnitude.

    The case study is the benchmark supercritical wing, using the transonic conditions from the AePW 1, 2 and 4. The results demonstrate the higher-order OSM-ROMs (greater than order two) are accurate for the prediction of forced unsteady aerodynamic response, flutter and LCO. Conversely, the ROMs based on first-order or second-order approximations are generally less accurate. A strong recommendation to extend this work would be to use a fixed linear model~\cite{levin22,brown22} based on strong theoretical foundation~\cite{silva97}, and add nonlinear terms using the method proposed in this paper. 

     Although this work makes a significant step towards applying this class of ROM to real-world aeroelastic problems the extension to a full aircraft model still comes with a significant computational burden considering that $i$) nonlinear generalized forces need to be identified for many structural modes, and $ii$) the CFD model would be significantly larger. For multi-input identification of a full aircraft, the number of nonlinear terms (including cross-terms) without sparsity to be identified would be in the order of hundreds-of-millions or billions. Significant opportunities exist in extension of this class of nonlinear ROM with sparsity promotion to full aircraft models.

     Using this optimization framework for the identification of nonlinear coefficients enables better comprehension of the nonlinear mechanisms at play in the system, see Appendix~\ref{sec:append}. Of particular interest to the BSCW would be to understand cross-coupling effects at $\alpha_0 = 5^\circ$ where the flutter mode changes from a coupled heave-pitch mechanism to a single-degree-of-freedom pitching mechanism. To properly assess cross-coupling effects, the heave mode ROM should be defined as a function of heave velocity rather than displacement. This was not conducted here given that the ROM performance was good without such a description. Furthermore, for larger systems (numerous structural modes), this method should provide a rapid means of assessing which modes are characterized by nonlinear cross-coupling and which are not.

    \section*{Acknowledgments}
    The authors are grateful for the ongoing financial support provided by the Australian Defence Science and Technology Group (DSTG). The authors are also grateful for the ongoing advice and support of Dr Walter Silva and Dr Pawel Chwalowski and from NASA Langley Research Center.

    \appendix
    \section{An Assessment of the Optimal Sparse Multi-Input Terms Identified}
    \label{sec:append}

    Table~\ref{tab:case2a_terms} summarizes the terms identified by OMP for the highest order ROM for each case. The largest reduction in the number of terms to be identified is for case 1, where the $5^{th}$-order ROM experiences a 99.95$\%$ reduction, while the reduction in the number of training samples of 96.88$\%$ (considering that 2500 training samples are used). Of the aeroelastic cases, the largest reduction in terms is 99.94$\%$ for case 3b, with a reduction in the number of training samples of 92.87$\%$ (considering that 3000 training samples are used). 
    
    It is of interest to inspect the quantity and order of cross-terms identified for each case to better understand the nonlinear cross-coupling effects in the system. It can be seen that for case 2a approximately half of the nonlinear terms are cross-terms. Of the cross-terms, a large percentage are second-order, suggesting that strong second-order cross-coupling exists a this condition. It is very interesting to observe that no nonlinear cross-terms are identified for case 2b. This may imply that at $M= 0.74$, as the AoA increases and the coupling mechanism between the pitch and heave mode weakens, there is less nonlinear cross-coupling between the modes. However, this observation does not hold for case 3 which could mean that at $M_\infty = 0.8$, where the nonlinear transonic mechanism is stronger, nonlinear cross-coupling is significant even at higher AoA. Although this analysis is incomplete, it highlights the benefits of the interpretability of the proposed ROM formulation. Recommendations to further this type of analysis are provided in the Conclusion.

    \begin{table}[h!]
    \centering
    \begin{tabular}{cccccccccccc}
        \hline
        && $1^{st}$-order & \multicolumn{2}{c}{$2^{nd}$-order} & \multicolumn{2}{c}{$3^{rd}$-order} & \multicolumn{2}{c}{$4^{th}$-order}& \multicolumn{2}{c}{$5^{th}$-order} & \multirow{ 2}{*}{total}\\
         &&  & direct & cross & direct& cross & direct& cross& direct& cross &  \\
        \hline
        \hline 
        \multirow{ 2}{*}{case 1} & $s_\alpha$ & 21 & 10 &-&8&-&4&-&5&-& 48 \\
         & $\kappa_\alpha$ & 22 &253&&2024&&12650&&65780&& 80279\\
         \hline
         \multirow{ 4}{*}{case 2a} & $s_h$ & 22 & 8 & 4 & 6 & 2 &&&&& 42 \\
         & $\kappa_h$ & 54 & 756 & 729 & 7308 & 20412 &-&-&-&-& 29259 \\
         & $s_\alpha$ & 10 & 4 & 5 &4 & 2 &-&-&-&-& 24 \\
         & $\kappa_\alpha$ & 14 & 56 & 49 & 168 & 392 &-&-&-&-&679 \\
         \hline
         \multirow{ 4}{*}{case 2b} & $s_h$ & 6 & 4 & 0 & 4 & 0 &-&-&-&-&14 \\
         & $\kappa_h$ & 54 & 756 & 729 & 7308 & 20412 &-&-&-&-&29259 \\
         & $s_\alpha$ & 5 & 3 & 0 & 2 & 0 &-&-&-&-&10 \\
         & $\kappa_\alpha$ & 50 & 654 & 621 & 5954 & 16146 &-&-&-&-& 23425\\
         \hline
         \multirow{ 4}{*}{case 3a} & $s_h$ & 7 &8&5&7&6&5&6&-&-&44 \\
         & $\kappa_h$ & 30 &240 &225&1360&3600&6120&34800&-&-&46375 \\
         & $s_\alpha$ & 5&7&3&2&1&8&0&-&-&26 \\
         & $\kappa_\alpha$ &28 &210&196&1120&2940&10880&26705&-&-&42079 \\
         \hline
         \multirow{ 4}{*}{case 3b} & $s_h$ &7&6&3&3&1&3&2&-&-&25 \\
         & $\kappa_h$ &28 &210&196&1120&2940&10880&26705&-&-&42079 \\
         & $s_\alpha$ &7&8&4&5&3&6&2&-&-&35 \\
         & $\kappa_\alpha$ &28 &210&196&1120&2940&10880&26705&-&-&42079 \\

        \hline
        
    \end{tabular}
    \caption{Linear and nonlinear terms identified for highest-order OSM-ROM for each case}
    \label{tab:case2a_terms}
    \end{table}

        \bibliographystyle{IEEEtranDOI}
        \bibliography{gensys_doi}

\end{document}